\documentclass[preprintnumbers,amsmath,amssymb,floatfix,10pt,prd,onecolumn,
superscriptaddress,nofootinbib]{revtex4}

\usepackage{latexsym}
\usepackage{epsfig}
\usepackage{epstopdf}
\usepackage{graphicx,float}
\usepackage{amssymb}
\usepackage{amsmath}
\usepackage{dcolumn}
\usepackage{bm}
\usepackage{color}
\usepackage{comment}

\usepackage[T1]{fontenc}
\usepackage[applemac]{inputenc}

\begin{document}
\title{\bf Thermodynamics and cosmological reconstruction in $f(T,B)$ gravity}

\author{Sebastian Bahamonde}
\email{sebastian.beltran.14@ucl.ac.uk}\affiliation{Department of Mathematics,University College London,
	Gower Street, London, WC1E 6BT, UK}

\author{M. Zubair}
\email{mzubairkk@gmail.com;drmzubair@ciitlahore.edu.pk}\affiliation{Department of Mathematics, COMSATS, Institute of Information
Technology Lahore,
	Pakistan}

\author{G. Abbas}
\email{abbasg91@yahoo.com}\affiliation{Department of Mathematics,
	The Islamia University of \\ Bahawalpur, Bahawalpur, Pakistan.}

\begin{abstract}
	Recently, it was formulated a teleparallel theory called $f(T,B)$  gravity which connects both $f(T)$ and $f(R)$ under suitable
        limits. In this theory, the function in the action is assumed to depend on the torsion scalar $T$ and also on a boundary term
        related with the divergence of torsion, $B=2\nabla_{\mu}T^{\mu}$. In this work, we study different features of a flat
        Friedmann-Lema\^{i}tre-Robertson-Walker (FLRW) cosmology in this theory. First, we show that the FLRW equations can be
        transformed to the form of Clausius relation
$\hat{T}_hS_{\rm eff}=-dE+WdV$, where $\hat{T}_h$ is the horizon temperature and
$S_{\rm eff}$ is the entropy which contains contributions both from
horizon entropy and an additional entropy term introduced due
to the non-equilibrium. 
We also formulate the constraint for the validity of the generalised second law of thermodynamics (GSLT).
        Additionally, using a cosmological reconstruction technique, we show that both $f(T,B)$ and $-T+F(B)$  gravity can mimic
        power-law, de-Sitter and $\Lambda$CDM models.
       Finally, we formulate the perturbed evolution equations and analyse the stability of some
important cosmological solutions.
\end{abstract}
\maketitle

\date{\today}

\section{Introduction}

In the current scenario, dark energy (DE) is referred as an active agent
which tends to accelerate the expansion in cosmos. The expanding
paradigm of the Universe has been affirmed from various
observational measurements. In 1998, observations of SNeIa
accumulated by the high-redshift SN team \cite{1} and SN cosmology
project team \cite{2} appearing as illuminating candles suggested an accelerating expansion of the Universe. The source for this
observed
cosmic
acceleration may be an anonymous energy component entitled as dark
energy. In spite of tremendous efforts, late-time cosmic
acceleration is certainly a major challenge for cosmologists. The
direct evidence for cosmic acceleration has strengthened over time
with measurements from temperature anisotropies in the cosmic microwave background (CMB) \cite{3} and
Baryon acoustic oscillations (BAO) \cite{4} which confirm the existence of DE. Dark energy is
appeared as an enigmatic cosmic ingredient and the interpretation of its
gravitational effects is a dynamic research field. The most likely
theoretical campaigner of DE is the cosmological constant $\Lambda$
characterized by a constant equation of state (EoS) $w=-1$
\cite{6}. A number of alternative models have been proposed in this
perspective to explain the role of DE in the present cosmic
acceleration \cite{4*}. The other proposal for the construction of
DE models is the modification of Einstein-Hilbert action which leads
to modified gravity models. Some important alternative theories of gravity  are $f(R)$ gravity \cite{12}, $f(R,\mathcal{T})$ gravity
($\mathcal{T}$ is the
trace of energy-momentum tensor $\mathcal{T}_{\alpha\beta}$) \cite{14},
$f(R,\mathcal{T},\mathcal{Q})$ gravity (where $\mathcal{Q}=R_{\alpha\beta}T^{\alpha\beta}$) \cite{15,15a},
Gauss-Bonnet gravity \cite{16}, teleparallel modifications \cite{13,13*,Bahamonde:2016kba}, scalar-tensor theories
\cite{17,Bahamonde:2015hza}, among others.

In current scenarios, generalization of teleparallel theory has
gained significant importance, which could provide alternative
explanations for the cosmic acceleration \cite{18}. A key problem in
$f(T)$ gravity is that it breaks the invariance under local Lorentz
transformations. Lack of local Lorentz symmetry implies that there
is no freedom to fix any of the components of the tetrad \cite{19}. Hence, loosing the Lorentz invariance means that two different
tetrads corresponding to the same metric could give different field equations. A new approach done in \cite{Krssak:2015oua}, introduced
a
new way to construct a covariant formulation of $f(T)$ gravity. Basically, in this approach the spin connection is chosen to be non-zero
and being pure-gauge. A more general theory containing the squares of the irreducible parts of torsion $f(T_{\rm ax},T_{\rm vec},T_{\rm
ten})$ has been also introduced in \cite{Bahamonde:2017wwk} using its covariant formulation. In our work, we will not use this covariant
approach, instead we will use the standard formulation of modified teleparallel theories of gravity where the spin connection is assumed
to be zero. In despite of the loose of the Lorentz invariance, this standard teleparallel approach has been very used in the literature.
One can somehow ``alleviative" the covariant issue (only at the level of the field equations) by choosing the correct tetrads
\cite{Tamanini:2012hg}. In FLRW cosmology, it is always possible to find ``good
tetrads" to obtain non-trivial cosmological solutions \cite{Tamanini:2012hg}. In flat FLRW cosmology, the diagonal tetrad in Cartesian
coordinates is a good tetrad since it does not restrict the field equations being the general relativity case as the tetrad in spherical
coordinates. In principle, at the level of the computations of the field equations, both approaches (covariant and non-covariant using
the good tetrads) should give the same result \cite{Bahamonde:2017wwk}. In \cite{13*},
the authors generalized $f(T)$ by introducing a new Lagrangian
$f(T,B)$ which involves a boundary term $B$ which is related to the
divergence of the torsion tensor. This theory becomes equivalent to
$f(R)$ gravity for the choice of special form $f(-T+B)$. The latter
is the only case in which Lorentz invariance can be achieved for a zero spin-connection. In our work, we are interested on studying
different cosmological properties of this theory as its thermodynamics laws, cosmological reconstruction method and the stability of
some
cosmological models.

Cosmological reconstruction is one of the most important tools that can be used in modified gravity to mimic realistic cosmological
scenarios. The reconstruction scheme in $f(R)$  gravity and its modifications have been
carried out under different scenarios \cite{z,z1,z2,z3} to find out
realistic cosmology which can explain the transition of matter dominated
epoch to DE phase. In this study, one interesting way is to consider the
known cosmic evolution and use the field equations to find particular form of
Lagrangian that can reproduce the given evolution background. Nojiri et al.
\cite{z3} executed such reconstruction scheme in order to find some realistic
models in $f(R)$ theory which was then applied in $f(R,\mathcal{G})$ modified
Gauss-Bonnet theories \cite{z4} (where $\mathcal{G}$ is the Gauss-Bonnet term). The cosmic evolution based on power law
solution of the scale factor has also been discussed in modified theories
\cite{z5}. Dunsby et al. \cite{z6} explored that extra degrees of freedom to
the matter component are necessary to reconstruct the $\Lambda$ cold dark
matter ($\Lambda$CDM) evolution in $f(R)$ gravity. Carloni et al. \cite{z7} set up a
new method of reconstructing $f(R)$ gravity using the cosmic parameters
rather than any form of the scale factor. In the context of modified
theories, stability of cosmological solutions has been analysed for
homogeneous perturbations \cite{z8,z9,Salako:2013gka,Setare:2013xh,Daouda:2011yf,z10}. In \cite{z9}, the stability
of $f(R,\mathcal{G})$ models is presented for power law and $\Lambda$CDM cosmology. In the context of teleparallel gravity, it was
showed
that one can reconstruct  $\Lambda$CDM universes and describe holographic dark energy models  for $f(T)$ gravity
\cite{Salako:2013gka,Setare:2013xh,Daouda:2011yf}. As $f(T,B)$ is a generalisation of $f(T)$ and $f(R)$ gravity, it is important to also
find out how this theory can reconstruct or mimic different cosmological models. One goal of this work is to reconstruct different
cosmological models for $f(T,B)$ gravity and also for the particular choice of the function where $f(T,B)=-T+F(B)$. After that, we will
also study the stability of some of these cosmological models.

The connection between the FLRW equations and the first law of thermodynamics (FLT) at the apparent
horizon was shown in \cite{30} for $\tilde{T}=1/2{\pi}R_A,~ S=\pi{R^2_A}/G$, where
$R_A$ is the radius of the apparent horizon and $\tilde{T}$ is the temperature. The Friedmann equations in
Gauss-Bonnet gravity and Lovelock gravity were also formulated by using the
corresponding entropy formula of static spherically symmetric black holes.
Eling et al. \cite{31} found that one cannot find the correct field equations
simply by using the Clausius relation in nonlinear theories of gravity. It
was remarked that a non-equilibrium treatment of thermodynamics is required,
whereby the Clausius relation is modified to $\tilde{T}dS=\delta{Q}+d_iS$, where
$d_iS$ is the entropy production term. Akbar and Cai \cite{32} showed that
the Friedmann equations in general relativity (GR) can be written as
$dE=\tilde{T}dS+WdV$ (unified FLT on the trapping horizon suggested by Hayward
\cite{33,34}) with the work term being $W=\frac{1}{2}(\rho-p)$. They also extended this
work to Gauss-Bonnet gravity \cite{32}, Lovelock gravity \cite{32,35},
braneworld gravity \cite{36,37}, $f(R)$ gravity \cite{39} and scalar-tensor
gravity \cite{40}. The generalised first and second laws of thermodynamics were also studied in the context of $f(T)$ gravity for
different forms of the function \cite{Karami:2012fu,Bamba:2012rv,Miao:2011ki}. As we have pointed out, the investigation about the
validity of thermodynamical laws in modified
theories has been carried out by numerous researchers in literature
\cite{41}. Here, we are interested to explore the validity of these laws in $f(T,B)$ gravity.\\
This paper is organised as follows: In Sec. II, we briefly introduce teleparallel equivalent of general relativity and then its
generalisation, $f(T,B)$ gravity. Then, we present the basis equations for a FLRW cosmology. Sec. III is devoted to the study of the
first and second laws of thermodynamics in this theory. Different reconstructions models are presented in Sec. IV for $f(T,B)$ and
also for $-T+F(B)$ gravity. Using perturbation techniques, the stability of different cosmological  models are studied in Sec. V.
Finally, in Sec. VI we conclude our main results.

\section{Teleparallel equivalent of general relativity and its modifications}
Let us briefly introduce the basis of the teleparallel equivalent of general relativity (TEGR). We will use the convention used in Ref.
\cite{13*} where $E^{\mu}_{m}$ is the inverse of the tetrad $e^{m}_{\mu}$ and Greek and Latin indices refer to space-time and tangent
space ones respectively. This theory lies in the idea that the manifold has a vanished curvature but a non-zero torsion. To ensure this
kind of geometry, one needs to chose a specific connection where the space is globally flat, the so-called Weitzenb\"{o}ck connection $
W_{\mu}{}^{a}{}_{\nu}$. One important fact it is that this alternative representation of gravity is equivalent (in the field equations)
to general relativity. The dynamical variable is the tetrad field and it is related with the metric with the following equation,
\begin{align}
g_{\mu\nu} &= e^{a}_{\mu} e^{b}_{\nu} \eta_{ab} \,, \label{metric}
\end{align}
where $\eta_{ab}$ represents the Minkowski metric $(-,+,+,+)$. Note that the tetrad fields are orthonormal vector at each point of the
manifold, hence they obey the following orthogonality relationships
\begin{align}
E_{m}^{\mu} e_{\mu}^{n} &= \delta^{n}_{m} \,,
\label{deltanm} \\
E_{m}^{\nu} e_{\mu}^{m} &= \delta^{\nu}_{\mu} \,.
\label{deltamunu}
\end{align}
The torsion tensor is constructed by taking the anti-symmetric part of the Weitzenb\"ock connection,
\begin{align}
T^{a}{}_{\mu\nu} &= W_{\mu}{}^{a}{}_{\nu} - W_{\nu}{}^{a}{}_{\mu} =
\partial_{\mu} e_{\nu}^{a} - \partial_{\nu}e_{\mu}^{a} \,.
\label{eq:tor}
\end{align}
The teleparallel action is then constructed with the torsion scalar which is defined as a contraction of the super potential
\begin{align}
S^{abc} = \frac{1}{4}(T^{abc}-T^{bac}-T^{cab})+\frac{1}{2}(\eta^{ac}T^b-\eta^{ab}T^c) \,,
\end{align}
with the torsion tensor $T=S_{a}{}^{bc}T^{a}{}_{bc}$. Here, the torsion vector is defined contracting the first two indices of the
torsion tensor $T_\mu=T^{\nu}{}_{\nu\mu}$. Explicitly, the action reads
\begin{align}\label{action1}
S_{\rm TEGR} =\frac{1}{\kappa^2} \int e\,T \, d^4x+ S_{\rm m} \,,
\end{align}
where $e$ denotes the determinant of the tetrad which is equal to $\sqrt{-g}$ and $\kappa^2=8\pi G$. Here, $S_{\rm m}$ is the action of the
matter content. It is possible to prove that the torsion scalar is related with the Ricci scalar directly by
\begin{align}
R= - T + \frac{2}{e}\partial_\mu (e T^\mu)=-T+B \,,
\label{ricciT}
\end{align}
where $B$ is a boundary term. The Einstein-Hilbert action is constructed with the Ricci scalar $R$, so that it differs only by a
boundary
term with respect to the TEGR action. Hence, tetrad variations of the action (\ref{action1}) are equivalent to metric variations of the
Einstein-Hilbert action. Therefore, if one varies the action (\ref{action1}) with respect to the tetrad, the corresponding field equations
will be identical to the Einstein field equations.\\
A well-studied modification of the action (\ref{action1}) is obtained by changing the torsion scalar $T$ to an arbitrary function $f(T)$
which depends smoothly on $T$. This generalisation then has the following action
\begin{align}
S_{f(T)} = \frac{1}{\kappa^2}\int e\,f(T)  \, d^4x+ S_{\rm m} \,,
\end{align}
which gives rise to the $f(T)$ field equations which is a second order theory. This theory is in this sense, analogous to  $f(R)$
gravity. However, these two theories are not equivalent. With the aim to combine both $f(R)$ gravity and $f(T)$ gravity, in Ref.
\cite{13*} it was proposed the following action
\begin{align}
\label{fTB}
S_{f(T,B)} =\frac{1}{\kappa^2} \int
dx^{4}\,e\,f(T,B) + S_{\rm m}\,,
\end{align}
which is a modified teleparallel theory of gravity where now $f(T,B)$ also depends on the boundary term $B$. In \cite{13*} it was proved
that by choosing $f=f(T)$ and $f=f(-T+B)=f(R)$ it is possible to recover both $f(T)$ and $f(R)$ gravity respectively. The field
equations
of this theory are obtained by varying the action with respect to the tetrad giving us,
\begin{multline}
2e\delta_{\nu}^{\lambda}\Box f_{B}-2e\nabla^{\lambda}\nabla_{\nu}f_{B}+
e B f_{B}\delta_{\nu}^{\lambda} +
4e\Big[(\partial_{\mu}f_{B})+(\partial_{\mu}f_{T})\Big]S_{\nu}{}^{\mu\lambda}
\\
+4e^{a}_{\nu}\partial_{\mu}(e S_{a}{}^{\mu\lambda})f_{T} -
4 e f_{T}T^{\sigma}{}_{\mu \nu}S_{\sigma}{}^{\lambda\mu} -
e f \delta_{\nu}^{\lambda} = 16\pi e\mathcal{T}_{\nu}^{\lambda} \,,
\label{fieldeq}
\end{multline}
where $\mathcal{T}_{\nu}^{\lambda}=e^{a}_{\nu}\mathcal{T}_{a}^{\lambda}$ is the standard energy momentum tensor and
$\Box=\nabla^{\mu}\nabla_{\mu}$. In general, this theory is a fourth-order one and in pure tetrad formalism, it is not invariant under local Lorentz
transformations (since $T$ and $B$ are not invariant under local LT). Indeed, the only theory which is invariant under these
transformations is obtained by taking $f(T,B)=f(-T+B)=f(R)$, i.e., in the $f(R)$ case.
\subsection{$f(T,B)$ Cosmology}
In this section, the basic equations for a flat FLRW cosmology in $f(T,B)$ will be introduced. The metric which describes this
space-time in Cartesian coordinates is given by
\begin{eqnarray}
ds^2&=&-dt^2+a(t)^2(dx^2+dy^2+dz^2)\,,
\end{eqnarray}
where $a(t)$ is the scale factor of the universe. In these coordinates, the tetrad field can be expressed as follows
\begin{equation}
e_{\mu }^{a}=\text{\textrm{diag}}\left( 1,a(t),a(t),a(t)\right)\,.
\label{fr.03}
\end{equation}%
If we assume that the content of the universe is
a perfect fluid and we use the above FLRW tetrad, the $f(T,B)$ cosmology field equations (\ref{fieldeq}) become
\begin{eqnarray}
-3H^2(3 f_{B}+2 f_{T})+3H \dot{f}_{B}-3\dot{H} f_{B}+\frac{1}{2} f(T,B)&=&\kappa^2 \rho_{\rm m}\,,\label{equation1}\\
-3H^2 (3f_{B}+2 f_{T})-\dot{H}(3 f_{B}+2 f_{T})-2H\dot{f}_{T}+\ddot{f}_{B}+\frac{1}{2} f(T,B)&=&-\kappa^2 p_{\rm m}\label{equation2}\,.
\end{eqnarray}
Here, $H=\dot{a}/a$ is the Hubble parameter and dots are differentiation with respect to $t$. Additionally, $\rho_{\rm m}$ and $p_{\rm m}$ are
the energy density and pressure of the matter content. It is easy to prove that the Ricci scalar is $R=-T+B=6(2H^2+\dot{H})$, where the
torsion scalar and the boundary term are $T=6H^2$ and $B=6(\dot{H}+3H^2)$ respectively. Moreover, by setting $f=f(T)$ or $f=f(-T+B)$ in
the above equations, we recover the standard $f(T)$ and $f(R)$ flat FLRW equations. One needs to be very careful with different metric
signature notations. In other $f(T)$ papers, some authors used a different signature notation where $\eta_{ab}=\textrm{diag}(1,-1,-1,-1)$
changing the sign of the torsion scalar $T\rightarrow -T=-6H^2$. For example, Eqs.~(2.9)-(2.10) reported in \cite{45} used
the other signature metric notation. To match those equations, one needs to change  $T\rightarrow -T=-6H^2$ and hence $f_T\rightarrow
-f_{T}$ and $f_{TT}\rightarrow f_{TT}$.

Eqs.~(\ref{equation1}) and (\ref{equation2}) can be
also represented in a fluid form,
\begin{eqnarray}\label{5*}
3H^2&=&\kappa^2_{\rm eff}\left({\rho}_{\rm m}+{{\rho}}_{TB}\right)\,,\\\label{6*}
2\dot{H}&=&-\kappa^2_{\rm eff}({\rho}_{\rm m}+p_{\rm m}+{\rho}_{TB}+{p}_{TB})\,.
\end{eqnarray}
The above equations are analogous to standard FLRW equations as in
GR, the quantities appearing in these equations are defined in terms
of $f(T,B)$ gravity as follows:
\begin{eqnarray}
\label{8*}
\rho_{\rm TB}&=&\frac{1}{\kappa^2}\Big[-3H\dot{f}_B+(3\dot{H}+9H^2)f_B-\frac{1}{2}f(T,B)
\Big] \,, \\\label{9*}
p_{\rm TB}&=&\frac{1}{\kappa^2}\Big[\frac{1}{2}f(T,B)+\dot{H}(2f_T-3f_B)-2H\dot{f}_T-9H^2f_B+\ddot{f}_B
\Big]\,,
\end{eqnarray}
where we have defined $\kappa^{2}_{\rm eff}$ as follows
\begin{eqnarray}\label{7*}
\kappa^2_{\rm eff}&=&-\frac{\kappa^2}{2f_T}\,.
\end{eqnarray}

Now, we have all the basis ingredients to study some properties in $f(T,B)$
cosmology as its thermodynamics and reconstructs some cosmological models.

\section{Thermodynamics of $f(T,B)$ gravity}
\subsection{Non-Equilibrium Description of Thermodynamics}
Here, we intend to explore the validity of thermodynamic laws in generalized
teleparallel theory of gravity in the non-equilibrium description. In the
following section, we determine the restriction on parameters and model of
$f(T,B)$ gravity for the validity of first and second laws of thermodynamics
at the apparent horizon of FLRW model. Also, we will show that for the total
energy of the system to be positive, it is necessary that graviton is not a
ghost in the sense of quantum gravity. We would like to mention that the
results of non-equilibrium thermodynamics in $f(R)$ and $f(T)$ theories can
be retrieved for some specific cases in this modified gravity.

The energy momentum tensor of additional geometric components satisfy the
following conservation equation
\begin{eqnarray}\label{7**}
\dot{\rho}_{\rm TB}+3H(\rho_{\rm TB}+p_{\rm TB})&=&\frac{T}{2\kappa^2}(2\dot{f}_T)\,,
\end{eqnarray}
Here, the energy conservation equation is not trivially satisfied since
$2\dot{f}_T\neq0$.

\subsubsection{First Law of Thermodynamics in $f(T,B)$ gravity}

In order to discuss the thermodynamics of $f(T,B)$ gravity, we can find the
dynamical apparent horizon by using the relation $h^{ab}{\partial}_a
\tilde{r}{\partial}_b \tilde{r}=0$. For the flat FLRW metric the radius of
the apparent horizon is
\begin{equation}\label{r}
\tilde{r}_A=\frac{1}{H}\,.
\end{equation}
The time derivative of the above equation gives us
\begin{equation}\label{dr}
-\frac{d{\tilde{r}_A}}{{\tilde{r}_A^3}}=\dot{H}Hdt\,.
\end{equation}
Using Eq.~(\ref{6*}) in the above equation, one gets
\begin{equation}\label{ddr}
\frac{f_Td{\tilde{r}_A}}{2\pi G}=-
{\tilde{r}_A}^3H({\rho}_{\rm eff}+{p_{\rm eff}})dt\,.
\end{equation}
Here, ${\rho}_{\rm eff}=\rho_{\rm TB}+\rho_{\rm m}$ and $p_{\rm eff}=p_{\rm TB}+p_{\rm m}$, are
the total density and pressure of the universe.

Now we need to define the Bekenstein-Hawking horizon entropy in
$f(T,B)$ gravity. For this purpose, we provide the review of such
definition in GR as well as in some non-standard theories. In GR,
Bekenstein-Hawking horizon entropy is defined by $S_h=A/(4G)$, where
$A=4\pi {\tilde{r}_A}^2$ is the area of the apparent horizon
\cite{42}. In modified theories of gravity like $f(R)$ gravity, the
horizon entropy $S_h$ associated with the Noether charge, the
so-called Wald entropy, can be defined by ${S}_h=A/(4G_{\rm eff})$
\cite{43}, where $G_{\rm eff}=G/f_R$ with $f_R=df(R)/dR$. We would
like to mention that this definition of Wald entropy in $f(R)$
gravity is valid for both metric and Palatini formulism \cite{44}. In
\cite{xi}, Brustein et al. showed that the Noether charge entropy is
equal to a quarter of the horizon area in units of the effective
gravitational coupling on the horizon defined by the coefficient of
the kinetic term of a specific metric perturbation polarization on
the horizon. They proposed that Wald's entropy can be expressed as
\begin{equation}\nonumber
S_h=\frac{A}{4G_{\rm eff}}\,.
\end{equation}
Similarly, in this notation, Wald entropy in $f(T)$ gravity is
defined as $S_h=2A/(4G_{\rm eff}),$ where $G_{\rm eff}=G/f_T$ \cite{45}. Hence in newly proposed modified
teleparalell gravity theory, we define the Wald entropy as
$S_h=A/(4G_{\rm eff}),$ where $G_{\rm eff}=-G/(2f_T)$. The Wald
entropy in $f(T,B)$ then reads as follows
\begin{equation}\label{S}
S_h=-\frac{A(2f_T)}{4G}\,.
\end{equation}
Clearly, if we set $f=f(-T+B)=f(R)$ and $f=f(T)$ we recover the standard $f(R)$ and $f(T)$ Wald entropy relationships respectively.
From Eqs.~(\ref{ddr}) and (\ref{S}), we get
\begin{equation}\label{dS}
-\frac{dS_h}{2\pi
\tilde{r}_A}=\frac{\tilde{r}_A}{G}df_T-4\pi{\tilde{r}_A}^3
H({\rho}_{\rm eff}+{p_{\rm eff}})dt\,.
\end{equation}

The associated temperature of the apparent horizon is defined
through the surface gravity $\kappa_{\rm sg}$ as
\begin{equation}\label{18*}
\tilde{T}_H=\frac{|\kappa_{\rm sg}|}{2\pi}\,,
\end{equation}
where $\kappa_{\rm sg}$ is given by \cite{17*}
\begin{eqnarray}\label{19*}
\kappa_{\rm sg}&=&\frac{1}{2\sqrt{-h}}\partial_{\alpha}(\sqrt{-h}h^
{\alpha\beta}\partial_{\beta}\tilde{r}_A)=-\frac{1}{\tilde{r}_A}
(1-\frac{\dot{\tilde{r}}_A}{2H\tilde{r}_A})=-\frac{\tilde{r}_A}{2}(2H^2+\dot{H})\,.
\end{eqnarray}
By multiplying the term
$\left(1-\frac{\dot{\tilde{r}}_A}{2H\tilde{r}_A}\right)$ on both sides of
Eq.~(\ref{dS}), we get
\begin{equation}\label{dSTH}
\tilde{T}_H{d{S}_h}=4\pi {\tilde{r}_A}^3
H({\rho}_{\rm eff}+{p_{\rm eff}})dt-2\pi{\tilde{r}_A}^2
({\rho}_{\rm eff}+{p_{\rm eff}})d\tilde{r}_A-\frac{\pi{\tilde{r}_A}^2\tilde{T}_H}{G}d(2f_T)\,.
\end{equation}
In GR, the Misner-Sharp energy is defined as $E=\tilde{r}_{A}/(2G)$.
For modified theories this relation is extended of the form
$E=\tilde{r}_{A}/(2G_{\rm eff})$. In $f(T,B)$ gravity, this definition
can be extended as
\begin{equation}\label{z}
E=-\frac{{\tilde{r}_A}(2f_T)}{2G}\,.
\end{equation}
From Eqs.(\ref{r}) and (\ref{z}), we then get that the Misner-Sharp energy is
\begin{equation}\label{E}
E=-V\frac{3H^2(2f_T)}{8\pi G}=V{\rho}_{\rm eff}\,,
\end{equation}
where $V=(4/3)\pi {\tilde{r}_A}^3$, is the volume of the interior region of
the apparent horizon. From the above equation, we find that $E$ is the total
intrinsic energy of the system. Also, we need to have $f_T<0$, so
that $E>0$. For this restriction on $f_T<0$, the effective
gravitational coupling $G_{\rm eff}=-G/(2f_T)$ needs to be positive. We
would like to mention that the condition $f_T<0$, is necessary
condition to ensure that graviton is not a ghost in the sense of quantum
gravity \cite{46}.

From Eqs.~(\ref{5*}) and (\ref{z}), one finds
\begin{equation}\label{zz}
dE=-\frac{\tilde{r}_A}{G}df_T+4\pi {\rho}_{\rm eff}
{\tilde{r}_A}^2d \tilde{r}_A-4\pi
H{\tilde{r}_A}^3({\rho}_{\rm eff}+{p_{\rm eff}})dt\,.
\end{equation}
By combining Eqs.~(\ref{dSTH}) and (\ref{zz}), one obtains
\begin{equation}\label{E11}
\tilde{T}_HdS_h=-dE+2\pi
{\tilde{r}_A}^2({\rho}_{\rm eff}-{p_{\rm eff}})d\tilde{r}_A-\frac{\tilde{r}_A}{G}(2\pi\tilde{r}_A\tilde{T}_H+1)df_T\,.
\end{equation}
By defining the work density, we get
\begin{equation}\label{E1}
W=-\frac{1}{2}\left(
\hat{T}{}^{(\rm M)\alpha\beta}h_{\alpha\beta}+\hat{T}{}^{(\rm DE)\alpha\beta}h_{\alpha\beta}\right)=\frac{1}{2}({\rho}_{\rm eff}-{p_{\rm eff}})\,.
\end{equation}
Here, $\hat{T}{}^{(\rm DE)\alpha\beta}h_{\alpha\beta}$ and $\hat{T}{}^{(\rm M)\alpha\beta}h_{\alpha\beta}$ are the energy-density of
the dark components and matter respectively.. Using the above definition of the work density in
Eq.~(\ref{E11}), one arrives at
\begin{equation}\label{E111}
\tilde{T}_HdS_h=-dE+WdV-\frac{\tilde{r}_A}{G}(2\pi\tilde{r}_A \tilde{T}_H+1)df_T\,,
\end{equation}
which can be re-written as
\begin{equation}\label{E4}
\tilde{T}_HdS_h+\tilde{T}_Hd\bar{S}=-dE+WdV\,,
\end{equation}
where $d\bar{S}=(\tilde{r}_A/(G\tilde{T}_H))(2\pi\tilde{r}_A\tilde{T}_H+1)df_T$.
The extra term $d\bar{S}$ defined in Eq.~(\ref{E4}) can be treated as an
entropy production term in non-equilibrium thermodynamics. Such additional
term marks to the non-equilibrium treatment of thermodynamics and is produced
internally due to the Lagrangian dependence both on the torsion scalar and
the boundary term.   The results in $f(R)$ and $f(T)$ theories can be
retrieved for some specific cases in this modified gravity. For the choice of
$f=f(T)$, we can reproduce the results of $f(T)$ gravity and the entropy
production term is the same reported in \cite{45}. Similarly for the choice of
$f(T,B)=f(-T+B)$, we find the results of $f(R)$ gravity \cite{BambaPLB}. In
literature, it has been shown that the theories involving non-minimal matter
geometry coupling also produce additional entropy production term (for review
see \cite{41}).

In $f(T)$ theory, the additional entropy term depends only on $T$, whereas in $f(T,B)$
gravity we have the contribution both from torsion and boundary terms. In $f(T,B)$ gravity $d\bar{S}\neq0$, due to $d(2f_T)\neq0$.
In GR and
alternative theories including Gauss-Bonnet and Lovelock gravities \cite{7*},
the usual FLT is satisfied by the respective field equations. In fact these
theories do not involve any surplus term in universal form of FLT
\emph{i.e.}, $\tilde{T}dS=-dE+WdV$. Here, we may define the effective entropy term
being the sum of horizon entropy and entropy production term as
$S_{\rm eff}=S_h+\bar{S}$ so that Eq.~(\ref{E4}) can be rewritten as
\begin{equation}
\tilde{T}_hdS_{\rm eff}=-dE+WdV\,,
\end{equation}
where $S_{\rm eff}$ is the effective entropy related to the contributions from
torsion scalar and boundary term at the apparent horizon of FLRW spacetime.

\subsubsection{Generalized Second Law of Thermodynamics}
In order to investigate the second law of thermodynamics in $f(T,B)$ gravity, one can start with the Gibbs equation in terms of matter and dark energy
components, given by
\begin{equation}\label{S1b}
\tilde{T}_{tot}dS_t=d(\rho_{\rm eff}V)+p_{\rm eff }
dV=Vd(\rho_{\rm eff})+({\rho}_{\rm eff}+p_{\rm eff})dV\,,
\end{equation}
where $S_t$ denotes the total entropy of the system inside the horizon.
It is natural to assume that the total temperature of energy source inside
the horizon is proportional to the temperature of the apparent horizon
\emph{i.e.}, $\tilde{T}_{\rm tot}=b\tilde{T}_h$ where $0<b<1$. It may result in local equilibrium by setting
the proportionality constant as unity as mentioned in \cite{39b***}.
Generically, the horizon temperature differs from the temperature of all
energy sources inside the horizon and the systems must experience interaction
for some interval of time ahead of attaining the thermal-equilibrium.
Furthermore, mutual coupling of matter and curvature components in this
theory may result in spontaneous energy flow between the horizon and matter
contents.

The validity of the generalised second law of thermodynamics (GSLT)
requires the condition
\begin{equation}\label{S1}
\Omega\equiv\frac{dS_h}{dt}+\frac{d(d\bar{S})}{dt}+\frac{d{S_t}}{dt}\geq0\,.
\end{equation}
Now using the FLRW equations together with Eqs.~(\ref{E4}) and (\ref{S1}), one
finds the following condition for validity of GSLT:
\begin{equation}\label{32*}
\frac{1}{2GH^4}\{(2-b)(2f_T)\dot{H}^2-2(1-b)(2f_T)\dot{H}H^2
-(1-b)H^3(2\dot{f}_T)\}\geq0\,.
\end{equation}
One can recover the expression of GSLT in $f(R)$ gravity presented in \cite{BambaPLB} under the transformation
$f(T,B)\rightarrow 2f(-T+B)=2f(R)$\footnote{We used the Lagrangian $ef(T,B)\kappa^2$, whereas in \cite{BambaPLB}
authors used the Lagrangian $\sqrt{-g}f(R)/(2\kappa^2)$}. Similarly, one can retrieve the results of $f(T)$ gravity if
$f(T,B)\rightarrow -2f(-T)$ \cite{45}. Note that in the later reference, the authors used the other signature notation for the metric so that the scalar torsion is equal to $T=-6H^2$ and not $T=6H^2$ as in our notation.

If $b=1$, i.e., temperature between outside and
inside the horizon remains the same then the GSLT is valid only if
\begin{align}\label{32**}
    \frac{3\dot{T}^2f_T}{2GT^3}\geq0\,.
\end{align}
For flat FLRW metric, one can define the effective components as
$\rho_{\rm eff}={\rho}_{\rm m}+{{\rho}}_{\rm TB}$ and $p_{\rm eff}=p_{\rm m}+{p}_{\rm TB}$ so that EoS is
defined as $w_{\rm eff}=-1-2\dot{H}/3H^2$. Here $\dot{H}<0$ corresponds to
quintessence region while $\dot{H}>0$ represents the phantom phase of the
universe. It follows that form \eqref{32**} that GSLT in $f(T,B)$ gravity is
satisfied in phantom era of cosmos. This result is compatible with \cite{46*}
according to which entropy may be positive even at the phantom era.

\subsection{Equilibrium Description of Thermodynamics}

In previous section, an additional entropy term $d\bar{S}$ in
formulation of laws of thermodynamics was found, which can be considered as the result
of non-equilibrium description of the field equations. In literature
\cite{Bamba:2012rv, 41, 44, 45}, it has been shown that the equilibrium
description does exist in modified theories of gravity and one can eliminate
the additional entropy term. Here, we discuss whether the equilibrium description of
thermodynamics in $f(T,B)$ gravity can be achieved or not.

We can rewrite the equations (\ref{5*})-(\ref{9*})
\begin{eqnarray}\label{e1}
3H^2&=&\kappa^2\left({\rho}_{\rm m}+{{\rho}}_{\rm TBE}\right)\,,\\\label{e2}
2\dot{H}&=&-\kappa^2({\rho}_{\rm m}+p_{\rm m}+{\rho}_{\rm TBE}+{p}_{\rm TBE})\,,
\end{eqnarray}
where now we have defined the new quantities as
\begin{eqnarray}\label{e3*}
\rho_{\rm TBE}&=&\frac{1}{\kappa^2}\Big[\frac{T}{2}(3f_B+2f_T+1)-3H\dot{f}_B+3\dot{H}f_B-\frac{1}{2}f(T,B)
\Big]\,, \\\label{e4*}
p_{\rm TBE}&=&\frac{1}{\kappa^2}\Big[\frac{1}{2}f(T,B)-\frac{T}{2}(3f_B+2f_T+1)-\dot{H}(3f_B+2f_T+2)-2H\dot{f}_T+\ddot{f}_B
\Big]\,.
\end{eqnarray}
The above equations are analogous to standard FLRW equations as in
GR plus the contribution of $f(T,B)$ gravity. Now we can check the validity of the first and second laws of
thermodynamics in this scenario.

\subsubsection{First Law of Thermodynamics}

In this representation of the field equations, the time derivative
of radius $\tilde{r}_A$ at the apparent horizon is given by
\begin{equation}\label{e5}
2d\tilde{r}_A=\kappa^2\tilde{r}^3_A({\rho}_{\rm m}+p_{\rm m}+{\rho}_{\rm TBE}+{p}_{\rm TBE})Hdt\,.
\end{equation}
Using the Bekenstein-Hawking entropy relation $S_{ h}=A/(4G)$, one gets
\begin{equation}\label{e6}
\frac{1}{2{\pi}\tilde{r}_A}dS_{h}=4{\pi}\tilde{r}^3_A({\rho}_{\rm m}+p_{\rm m}+{\rho}_{\rm TBE}+{p}_{\rm TBE})Hdt\,,
\end{equation}
and then by multiplying both sides of the above equation by $1-\dot{\tilde{r}}_{A}/(2 H
\tilde{r}_{A})$, implies that
\begin{equation}\label{e7}
\tilde{T}_{h} d\hat{S_{h}}=-4\pi\tilde{r}_{A}^{3}({\rho}_{\rm m}+p_{\rm m}+{\rho}_{\rm TBE}+{p}_{\rm TBE})H dt
+2\pi\tilde{r}_{A}^{2}({\rho}_{\rm m}+p_{\rm m}+{\rho}_{\rm TBE}+{p}_{\rm TBE})d\tilde{r}_{A}\,.
\end{equation}
Introducing the Misner-Sharp energy
\begin{equation}\label{e8}
\hat{E}=\frac{\tilde{r}_{A}}{2G}=V({\rho}_{\rm m}+{\rho}_{\rm TBE})\,,
\end{equation}
one obtains
\begin{equation}\label{e9}
dE=-4{\pi}\tilde{r}^3_A(\rho_{\rm m}+p_{\rm m}+{\rho}_{\rm TBE}+{p}_{\rm TBE})Hdt+4{\pi}
\tilde{r}^2_A(\rho_{\rm m}+p_{\rm m}+{\rho}_{\rm TBE}+{p}_{\rm TBE})d\tilde{r}_A\,.
\end{equation}
Now, by replacing Eq.~(\ref{e9}) into (\ref{e8}), one gets
\begin{equation}\label{e10}
\tilde{T}_{h}d\hat{S}_{h}=d\hat{E}-\hat{W}dV\,,
\end{equation}
where we have used the work density $\hat{W}=(1/2)(\hat{\rho}_{\rm eff}
-\hat{p}_{\rm eff})$ \cite{33,34}. Thus, as we have proved above, the equilibrium description of
thermodynamics can be derived by redefining the energy density $\rho_{\rm TB}$ and
the pressure $p_{\rm TB}$. Here, we find that the traditional first law of thermodynamics
$\tilde{T}_{h}d\hat{S}_{h}=d\hat{E}-\hat{W}dV$ can be met in equilibrium thermodynamic description of $f(T,B)$ gravity.
Hence, we can achieve first law of thermodynamics in similar fashion as in GR as well with modified theories including Gauss-Bonnet gravity \cite{32}, Lovelock gravity \cite{32,35} and braneworld gravity \cite{36,37}, $f(R)$ and $f(T)$ theories \cite{Bamba:2012rv,44, 45}.

\subsubsection{Generalized Second Law of Thermodynamics}

To establish the GSLT in this formulation of $f(T,B)$ gravity, one can
consider the Gibbs equation in terms of all matter field and energy
contents,
\begin{equation}\label{e11}
\tilde{T}_{\nu}dS_{\nu}=d(\rho_{\rm eff}V)+p_{\rm eff}dV\,,
\end{equation}
where $\tilde{T}_{\nu}$ denotes the temperature within the horizon. The second law of
thermodynamics can expressed as
\begin{equation}\label{e12}
\dot{S_{h}}+\dot{S}_{\nu}\geq0\,,
\end{equation}
where $S_{h}$, $S_{\nu}$ are the horizon entropy and the entropy
due to energy sources inside the horizon respectively. Now, we will assume a
relation between the temperature within the horizon and the temperature of
the apparent horizon given by
\begin{equation}\label{e13}
\tilde{T}_{\nu}=\tilde{T}_{h}\,.
\end{equation}
Substituting~(\ref{e10}) and (\ref{e11}) in (\ref{e12}), one obtains the condition for
the validity of GSLT which is
\begin{align}
    \frac{\dot{H}^2}{2GH^4}\geq0\,,
\end{align}
which shows that GSLT can be satisfied in equilibrium description of thermodynamics in similar pattern as in GR.
The reason behind the equilibrium description of thermodynamics is the validity of standard energy conservation
as compared to previous section of non-equilibrium thermodynamics. Moreover, in this scheme entropy is defined by the Bekenstein-Hawking entropy relation $S_{ h}=A/(4G)$, where entropy being proportional to horizon area. Here, we find that the second law of thermodynamics can be
met in both non-phantom and phantom phases as that in $f(R)$ and $f(T)$ theories \cite{BambaPLB,44,45}. It is remarked that this result is valid only if we have the same temperature of the universe outside and inside the apparent horizon \cite{zz}.

\section{Reconstruction method in $f(T,B)$ cosmology}\label{recons}
In this section, the usual reconstruction method will be used to find specific form of the function $f(T,B)$ which mimics different
cosmological models. Hereafter,
we will assume that the matter pressure is $p_{\rm m}=w \rho_{\rm m}$ where $w$ is the state parameter. Therefore, by using the matter conservation
equation, one finds
\begin{eqnarray}
\rho_{\rm m}(t)&=&\rho_{0}a(t)^{-3(w+1)}\,.\label{olaola}
\end{eqnarray}

\subsection{Power-law Cosmology}\label{Power-law}

It would be interesting to explore the existence of exact power
solutions in $f(T,B)$ gravity theory corresponding to different
phases of cosmic evolution.  Let us consider a model described by a power-law scale factor given by
\begin{eqnarray}
a(t)&=&\Big(\frac{t}{t_{0}}\Big)^{h}\,,\label{42}
\end{eqnarray}
where $t_{0}$ is some fiducial time and $h$ is greater than zero. These solutions help to explain the
cosmic history including matter/radiation and dark energy dominated
eras. Further, these solutions provide the scale factor evolution
for the standard fluids such as dust ($h=2/3$) or
radiation ($h=1/2$) dominated eras of the Universe. Also,
${h>1}$ predicts a late-time accelerating Universe. We would like
to mention that $h$ is arbitrary constant in the following form of
power law solutions. For the above scale factor, the scalar torsion and
boundary read as follows
\begin{eqnarray}
T&=&\frac{6h^2}{t^2}\,,\label{TT1}\\
B&=&\frac{6 h (3 h-1)}{t^2}\,.\label{BB1}
\end{eqnarray}
Now, for simplicity, we will assume that the function can be written in the following form
\begin{eqnarray}
f(T,B)&=&f_{1}(T)+f_{2}(B)\,.\label{sep}
\end{eqnarray}
By inverting (\ref{TT1}) and (\ref{BB1}), the $00$ equation given by (\ref{equation1}) becomes
\begin{align}
\frac{1}{2}f_{1}(T)-T f_{1,T}(T)-\kappa^2 \rho_{\rm m}(t)&=K\,,\\
-2 B^2 f_{2,BB}(B)+(1-3h) B f_{2,B}(B)+(3 h-1) f_{2}(B)&=(2-6 h)K\,.
\end{align}
Here, $K$ is a constant for the method of separation variable and $f_{1,T}=df_1/dT$ and $f_{2,B}=df_2/dB$. We can directly solve the above equations obtaining
\begin{eqnarray}\label{pfTB}
f_{1}(T)&=& \frac{2\kappa^2 \rho_{0}}{1-3 h (w+1)}\Big(\frac{t_{0}}{\sqrt{6}h}\sqrt{T}\Big)^{3h(w+1)}+C_1 \sqrt{T}+2K\,,\\\label{pfTB1}
f_{2}(B)&=&C_{2}B^{\frac{1}{2} (1-3h)}+C_{3}B-2K\,.
\end{eqnarray}
Note that this is one specific form of the function which mimics a power-law cosmology. There are other possible functions that also
will
represent this model. The separation of variable can be done either by choosing that $\rho_{\rm m }$ depends on $T$ or $B$. The later comes
from
the fact that $\rho_{\rm m}=\rho_{\rm m}(t)$ and also $T=T(t)$ an $B=B(t)$. Hence, in principle, the energy density is $\rho(t)=\rho_0
(\frac{t}{t_0})^{-3h(w+1)}$ and by using (\ref{TT1}) and (\ref{BB1}) one can rewrite the energy density in two ways, namely
\begin{eqnarray}
\rho_{\rm m}(T)&=&\rho_0 \left(6^{h/2} \left(\frac{h}{\sqrt{T} t_0}\right)^h\right)^{-3 (w+1)}\,,
\end{eqnarray}
or
\begin{eqnarray}
\rho_{\rm m}(B)&=&\rho_0 \left(6^{h/2} \left(\frac{\sqrt{h (3 h-1)}}{\sqrt{B} t_0}\right)^h\right)^{-3 (w+1)}\,.
\end{eqnarray}
Hence, one has a freedom to choice either $\rho_{\rm m}=\rho_{\rm m}(T)$  or $\rho_{\rm m}=\rho_{\rm m}(B)$ in the separation of variables. In our case, we
chose
$\rho_{\rm m}=\rho_{\rm m}(T)$ but in principle, another kind of solution for the reconstruction method can be found by choosing  $\rho_{\rm m}=\rho_{\rm m}(B)$. A
similar approach was done in Sec. 5.1 in \cite{delaCruzDombriz:2011wn} and also in \cite{delaCruz-Dombriz:2017lvj}. As it is stated in those references, in TEGR the density matter is usually described by $T$ so that without loosing any generality, we will also use that approach. Then, for the following sections, we will use that
$\rho_{\rm m}=\rho_{\rm m}(T)$ instead of $\rho_{\rm m}=\rho_{\rm m}(B)$. Note that from Eqs.~(\ref{TT1}) and (\ref{BB1}), one can also express $T=T(B)$ or $B=B(T)$. In principle, one can try to solve the 00 equation (\ref{equation1}) just by changing all in terms of $T$. However, this procedure makes the equation very complicated and it is almost impossible to find an analytical solution for the function $f$. As pointed out before, this kind of behaviour is something well-known in reconstruction techniques when one is considering two functions in $f$. See for example \cite{delaCruzDombriz:2011wn} and also \cite{delaCruz-Dombriz:2017lvj} where they also can express either $T=T(T_G)$ or $R=R(G)$ in their theories. In those papers, one can also see the situation described above.

\begin{figure}[H]
\center\epsfig{file=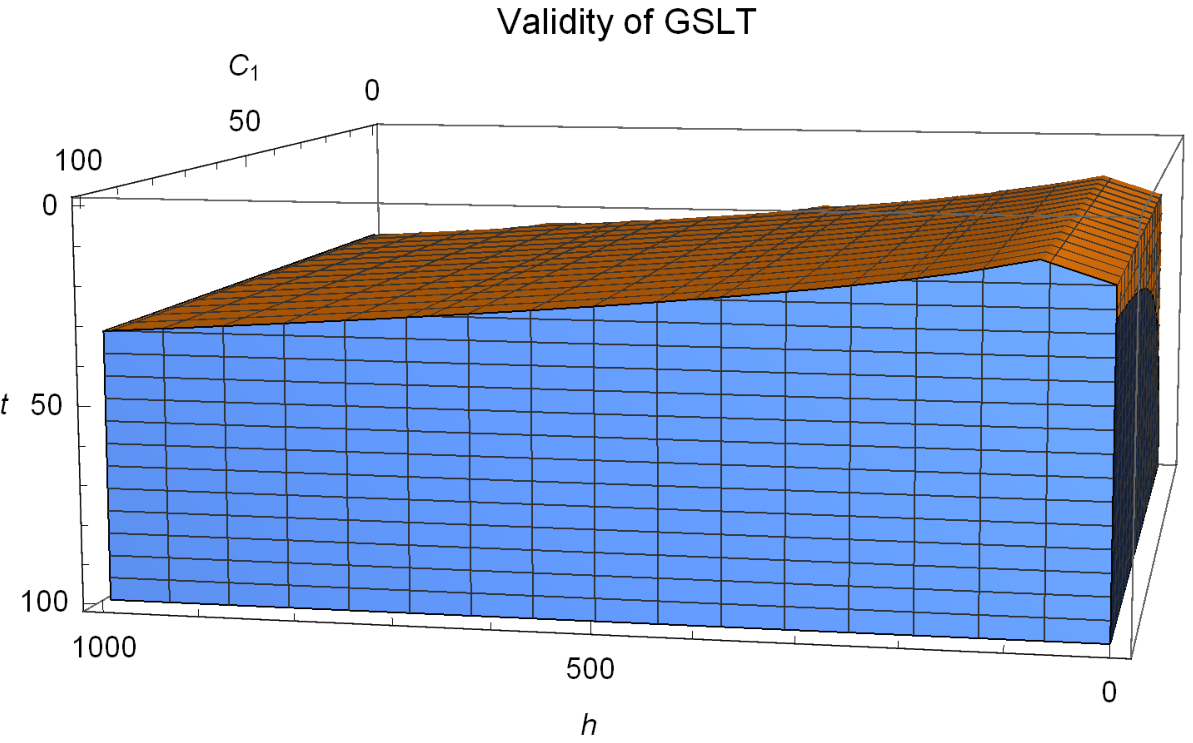, width=0.45\linewidth}\epsfig{file=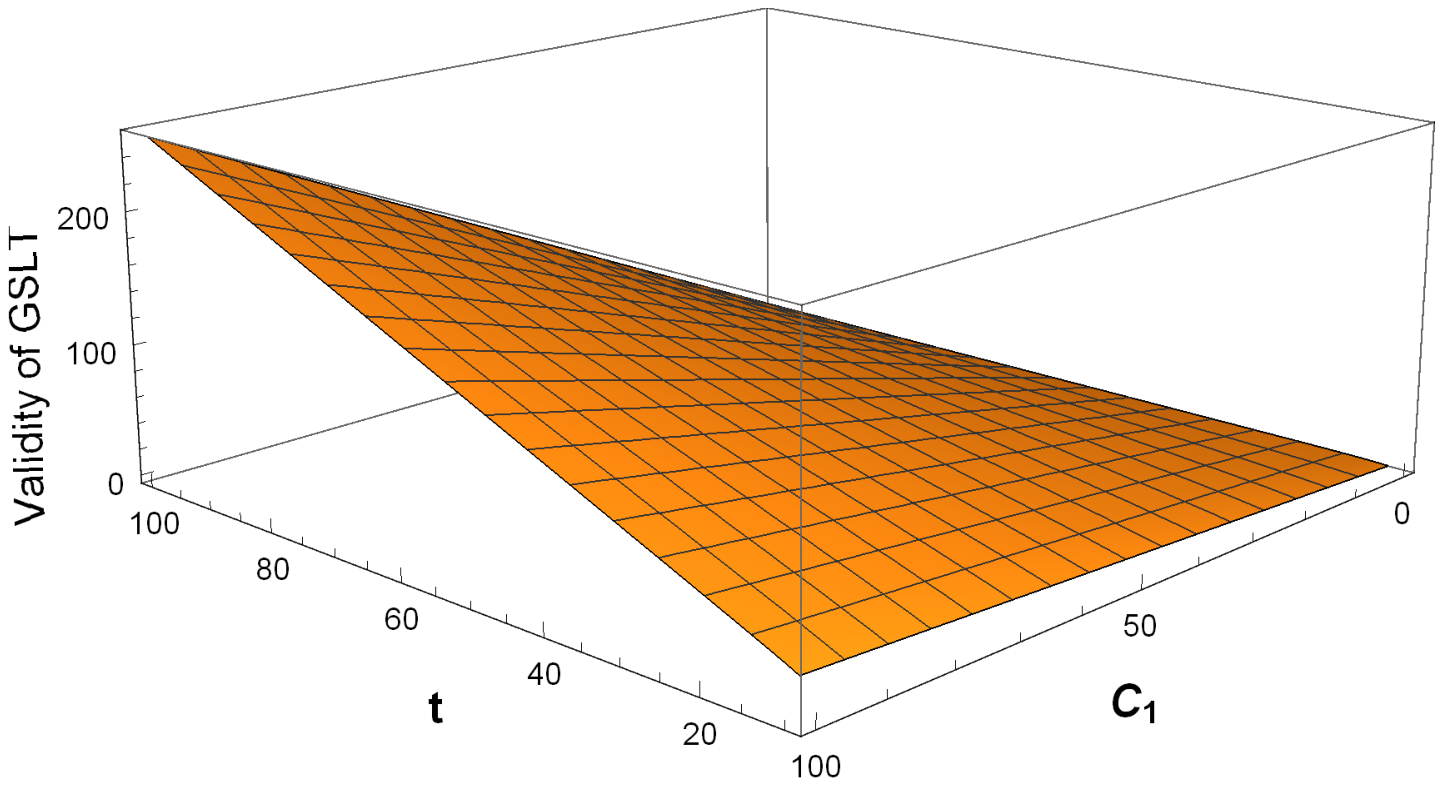, width=0.46\linewidth}
\caption{The figure on the left represents the regions where GSLT is valid for $h>1$ and $C_1\geq0$ whereas
the figure on the right shows behavior of GSLT for $h=2$. Herein, we set $w=0$ and $H_0=67.3$. }
\label{fig11}
\end{figure}

Now we explore the validity of GSLT for the power-law $f(T,B)$ model. By substituting Eqs.~(\ref{pfTB}) and (\ref{pfTB1}) into (\ref{32**}), one finds the following result for the validity of GSLT:
\begin{eqnarray}
\frac{\dot{H}^2}{2TH^4}\left(C_1\sqrt{T}+\frac{6^{1-\frac{3}{2}h(1+w)}\rho_0h\kappa^2(1+w)
\left(\frac{\sqrt{T}t_0}{\sqrt{h}}\right)^{3h(1+w)}}{1-3h(1+w)}\right)\geq0\,.
\end{eqnarray}
Here, the validity of GSLT depends on the constant $C_1$ and the power-law parameter $h$. On the left of Fig.~\ref{fig11}, the evolution of GSLT is depicted by varying both $C_1$ and $h$. For $C_1\geq0$, it is found that the validity period decreases when $h$ increases. In the plot on the right in Fig.~\ref{fig11}, we choose a particular value of $h$ to show the validity of GSLT.

\subsection{de-Sitter reconstruction}

If one assumes that the universe is governed by a de-Sitter form, i.e., the
scale factor of the universe is an exponential $a(t)\propto e^{H_{0}t}$, both
the torsion scalar and the boundary term are constants. Explicitly they are
given by $T=6H_{0}^2$ and $B=18H_{0}^2$ respectively. This kind of evolution
of the universe is very well known and important since it correctly describes
the expansion of the current universe. In GR, for this kind
of universe, is known that the universe must be filled by a dark energy fluid
whose state parameter $w=-1$ and hence the energy density is also a constant.
From our modified theory, a priori $w=-1$ does not need to describe De-Sitter
universes. Hence, to find de-Sitter reconstruction we must set $H=H_0$. From Eq. (\ref{equation1}), it is easily to see that any kind of
functions of $f(T,B)$ can admit de-Sitter solution if the following
constraint is satisfied,
\begin{eqnarray}\label{50}
H_{0}^2\left(9 f_{B}(T_{0},B_{0})+6f_{T}(T_{0},B_{0})\right)-\frac{1}{2} f(T_{0},B_{0})=0\,.
\end{eqnarray}
For instance, by assuming that the function is separable as $f(T,B)=f_{1}(B)+f_{2}(T)$, a possible reconstruction function which
describes a de-Sitter universe is given by
\begin{eqnarray}\label{dS1}
f(T,B)&=&2 (\kappa ^2 \rho_0+2 K)+f_{0}e^{\frac{B}{18 H_{0}^2}}+\tilde{f}_{0}e^{\frac{T}{12 H_{0}^2}}\,,
\end{eqnarray}
which of course is a constant function. Here, $f_{0}$ and $\tilde f_{0}$ are integration constants. In case of de-Sitter model, it can
found that GSLT is trivially satisfied.

\subsection{$\Lambda$CDM reconstruction}

Here, the reconstruction of the $f(T,B)$
function for a $\Lambda$CDM cosmological evolution will be discussed in the absence of any
cosmological constant term in the modified Einstein field equations.
This model was firstly formulated by Elizalde et al.~\cite{CDM} in
$f(R,\mathcal{G})$ modified theory of gravity. The cosmological effects of
the cosmological constant term in the concordance model is exactly
replaced by the modification introduced by $f(T,B)$ function with
respect to the usual Einstein-Hilbert Lagrangian.

For simplicity, instead of working with all the variables depending
on the cosmic time $t$, the e-folding parameter defined
as $N=\ln\left(a/a_0\right)$ will be used. By using $a(t)=a_{0}/(1+z)$, the e-folding parameter can be also written depending on the redshift
function
$z$  as $N=-\ln(1+z)=\ln(1/(1+z))$. In terms of this variable,
one can express $a(t),~H(t)$ and time derivatives as
\begin{equation*}
a=a_0e^N, \quad H=\frac{\dot{a}}{a}=\frac{dN}{dt}, \quad
\frac{d}{dt}=H\frac{d}{dN}\,.
\end{equation*}
Therefore, one can rewrite equation (\ref{equation1}) in terms of $N$, yielding
\begin{equation}\label{fTB11}
-3H^2(3 f_{B}+2 f_{T})+18H\Big[(H^2H''+HH'^2+6H^2H')f_{BB}+2H^2H'f_{BT}\Big]
-3HH' f_{B}+\frac{1}{2} f(T,B)=\kappa^2\rho_{\rm m}(t)\,.
\end{equation}
Here, primes denote differentiation with respect to the e-folding $N$. Additionally, in term of the e-folding, the scalar torsion and
the
boundary term are $T=6H^2$ and $B=6H(3H+H')$ respectively.
Now, for convenience, we introduce a new variable $g=H^2$ making that the above equation becomes
\begin{eqnarray}
-\frac{3}{2} \left(g'+6 g\right) f_{B}+18 g g' f_{TB}+9 g f_{BB} \left(g''+6 g'\right)-6 g f_{T}+\frac{1}{2}
f(T,B)=\kappa^2\rho_{\rm m}(t)\,.
\end{eqnarray}
It is easily to compute that the torsion scalar and the boundary term written in this variable are $T=6 g$ and $B=3 (g'+6 g)$
respectively. Now, it will be also assumed that the function $f(T,B)$ is separable as Eq.~(\ref{sep}). Using these assumptions, the above
equation
becomes
\begin{eqnarray}
-\frac{3}{2} \left(g'+6 g\right) f_{1,B}(B)+9 g f_{1,BB}(B) \left(g''+6 g'\right)+\frac{1}{2}f_{1}(B)=\nonumber\\
\kappa^2\rho_{\rm m}(t)-\frac{1}{2}f_{2}(T)+6 g f_{2,T}(T)\,.\label{eqq}
\end{eqnarray}
Let us now reconstruct the $\Lambda$CDM model whose function $g=g(N)$ is given by \cite{CDM}
\begin{align}
g&=H_{0}^2+le^{-3N}, \ \ \ l=\frac{\kappa^2\rho_{0}a_{0}^{-3}}{3}\,.
\end{align}
In this model, the e-folding can be expressed depending on the boundary term and the torsion scalar as follows
\begin{eqnarray}
N=\frac{1}{3} \log \left(\frac{9 l}{B-18 H_{0}^2}\right)=\frac{1}{3} \log \left(-\frac{6 l}{6 H_0^2-T}\right)\,.\label{NN}
\end{eqnarray}
Therefore, one can rewrite Eq.~(\ref{eqq}) as follows
\begin{eqnarray}
2 \left(27 B H_0^2-162 H_0^4-B^2\right)f_{1,BB}(B)-B f_{1,B}(B)+f_{1}(B)&=&K\,,\\
\kappa^2 \rho_{0}  \left(\frac{6 H_{0}^2-T}{-6la_{0}^3}\right)^{w+1}-\frac{1}{2}f_{2}(T)+T f_{2,T}(T)&=&\frac{K}{2}\,,
\end{eqnarray}
where $K$ is a constant since the r.h.s. of  (\ref{eqq}) depends only on $T$ and the l.h.s. only on $B$. Note that the energy density
can
be expressed depending on $T$ or $B$ so that, the above equations are one of the possible options to reconstruct a $\Lambda$CDM
Universe. Thus, by solving the above equations, one gets that one way to reconstruct $\Lambda$CDM  is by taking the following functions,
\begin{eqnarray}\label{z*}
f_{1}(B)&=&\frac{C_2 \left(3 H_0 \sqrt{B-9H_0^2}-B \arctan\left(\frac{\sqrt{B-9 H_{0}^2}}{3 H_{0}}\right)\right)}{54 H_{0}^3}+B
C_1+K\,,\\\label{z**}
f_{2}(T)&=&K+C_{3}\sqrt{T} +\frac{\kappa ^2 \rho_{0}}{3}H_{0}^{2w}(a_{0}^3 l)^{-(w+1)}\left(6 H_{0}^2 \,
_2F_1\left(-\frac{1}{2},-w;\frac{1}{2};\frac{T}{6 H_{0}^2}\right)+T \, _2F_1\left(\frac{1}{2},-w;\frac{3}{2};\frac{T}{6
H_{0}^2}\right)\right)\,,
\end{eqnarray}
where $H_0\neq0$ and for the case where $H_0=0$ we find
\begin{eqnarray}\label{rec1}
f_{1}(B)&=&B C_1+\frac{C_2}{\sqrt{B}}+K\,,\\
f_{2}(T)&=&-\frac{2^{-w} \kappa ^2 \rho_0 }{2 w+1}\left(\frac{T}{3 a_{0}^3 l}\right)^{w+1}+C_3 \sqrt{T}-K\,.
\end{eqnarray}
Here, $C_{1}$, $C_{2}$ and $C_{3}$ are constants and $_2F_1$ represents the hypergeometric function of the second kind. The case where
$H_0=0$ represents a power-law solution with $h=2/3$. The above solution is consistent  with Eqs.~(\ref{pfTB}) and (\ref{pfTB1})
in that limit. Note that the case $T=6H_0^2,B=18H_0^2$ which represents de-Sitter universes can not be recovered directly from the above
equations. However, these models can be recovered directly from (\ref{eqq}) by imposing $g=H_0^2$ (with $l=0$) which actually gives us
the same result obtained in the previous section (see Eq. (\ref{dS1})). This issue comes from the fact that one expresses the e-folding
depending on $B$ or $T$, one needs to assume $l\neq0$. The same issue can be seen in Sec.~2.1 in \cite{z4}.

\begin{figure}[H]
\center\epsfig{file=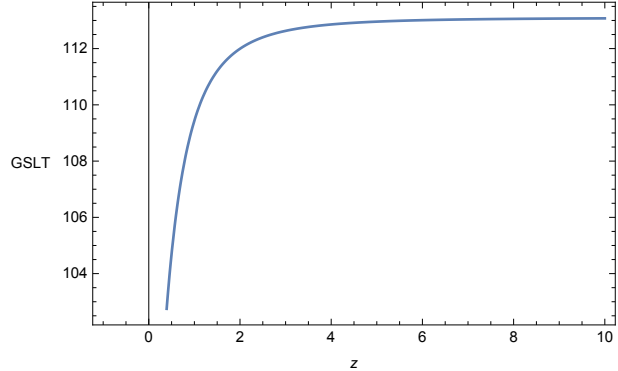, width=0.45\linewidth}\caption{Plot of GSLT (defined in Eq.~\eqref{ddd}) versus the redshift function $z$ for
$\Lambda$CDM model. Here, $C_3=0.1$ is assumed.}
\label{fig22}
\end{figure}

For the above model (\ref{z*})-(\ref{z**}), the validity constraint for GSLT takes the following form
\begin{eqnarray}\label{ddd}
&&\frac{3(g-H_0^2)^2}{8g^2(la_0^3)^{1+w}}\left((la_0^3)^{1+w}\frac{\sqrt{6g}}{C_3}+2H_0^{2w}\kappa^2
\left[\left(1-\frac{g^2}{H_0^2}\right)^w\right.\right.\\\nonumber\times&&\left.\left.(g^2-H_0^2)+H_0^2
2F_1\left(-\frac{1}{2},-w;\frac{1}{2};\frac{g^2}{H_{0}^2}\right)+g 2F_1
\left(\frac{1}{2},-w;\frac{3}{2};\frac{g^2}{H_{0}^2}\right)\right]\rho_0\right)\geq0\,.
\end{eqnarray}
Fig.~\ref{fig22} shows the validity of GSLT (above equation) as a function of redshift $z$ for the specific case where $C_3=0.1$. It can be seen that for this case, expression~\eqref{ddd} is always positive ensuring that GSLT is valid for any redshift.
(see Figure 2).

\subsection{Phantom behaviour}\label{secphantom}

\begin{figure}[H]
\center\epsfig{file=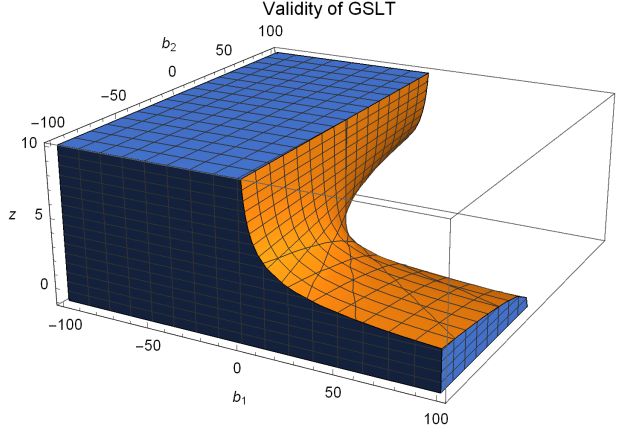, width=0.45\linewidth}\caption{Plot of the validity of GSLT for the phantom dominated model (defined in \eqref{ols}). Here, $h_0=0.1$, $m=1$, and $C_3=1$ are set.}\label{fig33}
\end{figure}

As an another example, let us now assume that the Hubble parameter and the energy density are \cite{CDM}
\begin{eqnarray}\label{phantom}
\sqrt{g}=H=h_{0}e^{mN}\,, \  \rho_{\rm m}=b_{0}+b_{1}e^{2mN}+\frac{96(m+1)}{5}b_{2}e^{5mN}\,,
\end{eqnarray}
where $h_{0}$, $b_{0}$, $b_{1}$, $b_{2}$ and $m$ are constants. Cosmologically speaking, this model represents a super accelerated
universe phase with a phantom regime $w_{\rm eff}<-1$, making that the universe could end in a singularity. If one uses Eq~ (\ref{eqq}), one can split the cosmological equations depending only on $T$ and $B$ as follows ($m\neq -3$)
\begin{eqnarray}
2 B^2 m f_{1,BB}(B)-B (m+3) f_{1,B}(B)+(m+3) f_1(B)&=&K (m+3)\,,\\
\kappa ^2 \left(b_{0}+\frac{b_{1} T}{6 h_{0}^2}+\frac{4}{15} \sqrt{\frac{2}{3}} b_{2} (m+1)
\left(\frac{T}{h_{0}^2}\right)^{5/2}\right)+T
f_{2,T}(T)-\frac{f_{2}(T)}{2}&=&\frac{K}{2}\,,
\end{eqnarray}
where $K$ is a constant. Thus, one of the possible representation which produces a super accelerated universe is given by the following
functions,
\begin{eqnarray}
f_{1}(B)&=&K+C_{1}B^{\frac{3+m}{2m}}+C_{2}B\,,\\
f_{2}(T)&=&2 b_{0} \kappa ^2-K+C_3 \sqrt{T}-\frac{b_{1}\kappa ^2 T}{3 h_{0}^2}-\frac{2 \sqrt{\frac{2}{3}} b_{2} \kappa ^2 (m+1)
T^{5/2}}{15 h_{0}^5}\,,
\end{eqnarray}
where $C_{1}$, $C_{2}$ and $C_{3}$ are integration constants.
For the above model, let us now explore the constraint for the validity of GSLT which becomes of the form
\begin{eqnarray}\label{ols}
\frac{m^2}{2}\left(-\frac{2\kappa^2b_1}{3h_0^2}-
\frac{8(1+m)\kappa^2b_2g^3}{h_0^8}+\frac{C_3}{\sqrt{6g}}\right)\geq0\,.
\end{eqnarray}
One can notice that GSLT is always satisfied for any values of $C_3$ and $h_0$ if both $b_1$ and $b_2$
are assigned negative values. As an example, in Fig.~\ref{fig33}, the validity region of GSLT is depicted for the parameters $b_1,b_2$ and $z$, where it was set $h_0=0.1$, $m=1$, and $C_3=1$.

\subsection{Reconstruction method in $f(T,B)=-T+F(B)$ cosmology}

 In this section, the specific case where the function takes the form $f(T,B)=-T+F(B)$ will be studied, which is similar to models of the
 form $f(R)=R+F(R)$ and $f(T)=-T+f(T)$ studied in $f(R)$ and $f(T)$ gravity respectively \cite{various1}. This theory is equivalent  to consider a teleparallel background (or GR) plus an additional function which depends on the boundary term which can be also
 understood as $F(B)=F(T+R)$. It is important to mention that even though the case $f(T,B)=f_{1}(B)+f_{2}(T)$ studied in the previous
 section is more general and in principle could contain the case  $f(T)=-T+F(B)$, one might get a different reconstruction solution. The
 later comes from the fact that the case $f(T)=-T+F(B)$ is a very specific choice of the function and also that all the functions found
 before in Sec.~\ref{recons} are one of the possible choices for reconstructing the corresponding models. Moreover, due to the
 mathematics technique employed before, i.e., the method of separation of variables, if one tries to recover the case  $f(T)=-T+F(B)$ from
 the solution, one might not get the same answer. As an example, for the power-law case is not possible to recover $f(T,B)=-T+F(B)$
 unless
 we restrict our model with $C_{1}=0$ and $h=\frac{2}{3(w+1)}$ which is only a kind of power-law model (see Eqs.~(\ref{pfTB}) and
 (\ref{pfTB1})). Hence, it is interesting and important to also study if it is possible to reconstruct these cosmological models within
 this particular theory.\\
 In this model, the $00$ field equation (\ref{equation1}) becomes
\begin{eqnarray}
-3H^2(3 F_{B}-2)+3H \dot{F}_{B}-3\dot{H} F_{B}+\frac{1}{2} (F(B)-6H^2)&=& \kappa^2\rho_{\rm m}(t)\,,\label{equation1pri}
\end{eqnarray}
where the energy density is given by (\ref{olaola}). Equivalently, from (\ref{fTB11}), it is easily to rewrite the above equation in
term
of the e-folding,
\begin{equation}\label{fTB111}
-3H^2(3 F_{B}-2)+18H\Big[(H^2H''+HH'^2+6H^2H')F_{BB}\Big]
-3HH' F_{B}+\frac{1}{2} (-6H^2+F(B))=\kappa^2\rho_{\rm m}(t)\,.
\end{equation}
Let us now perform a reconstruction method for all the same models studied in Secs.~\ref{Power-law}-\ref{secphantom}. \\
For a power-law cosmology described in Sec.~\ref{Power-law}, Eq.~(\ref{equation1pri}) can be written as follows,
\begin{eqnarray}
\frac{B \left(h-2 B F_{BB}(B)\right)}{6 h-2}-\frac{1}{2} B F_B(B)+\frac{F(B)}{2}=\kappa^2\rho_{0}\left(\frac{6 h (3 h-1)}{B
t_0^2}\right)^{-\frac{3}{2} h (w+1)}\,,
\end{eqnarray}
which can be directly solved, yielding the following solution
\begin{eqnarray}
F(B)&=&C_1 B^{\frac{1-3h}{2}}+B \left(C_2-\frac{2 h}{(3 h+1)^2}\right)+\frac{h B\log (B)}{3 h+1}+\frac{B h \log (9 h+3)}{3
h+1}\nonumber\\
&&-\frac{\kappa ^2 \rho_{0} \left((3 h-1)^{1-\frac{3}{2} h (w+1)} 2^{2-\frac{3}{2} h (w+1)}\right) \left(\frac{B t_{0}^2}{3
h}\right)^{\frac{3}{2} h (w+1)}}{(3 h (w+1)-2) (3 h (w+2)-1)}\,,
\end{eqnarray}
where $C_{1}$ and $C_{2}$ are integration constants.\\
Now, for a de-Sitter reconstruction, the scale factor behaves as $a(t)=a_{0}e^{H_{0}t}$, then $B=18H_{0}^2$ and hence from
(\ref{equation1pri}) we directly find that the function takes the following form,
\begin{eqnarray}
F(B)&=& C_1 e^{\frac{B}{18H_0^2}}-2 \left(3H_0^2-\kappa ^2 \rho_0\right)\,.
\end{eqnarray}
Here, $C_{1}$ is an integration constant. Let us now reconstruct a $\Lambda$CDM universe where $g=H_{0}^2+le^{-3N}$. In this theory,
Eq.
(\ref{eqq}) becomes
\begin{eqnarray}
-\left(B-18 H_{0}^2\right) \left(B-9 H_{0}^2\right) F_{BB}(B)-\frac{1}{2} B F_{B}(B)+\frac{F(B)}{2}+\frac{1}{3} \left(B-9
H_{0}^2\right)=\nonumber\\
3^{-2 (w+1)}\kappa^2 \rho_{0}\left(a_{0} \sqrt[3]{\frac{l}{B-18 H_{0}^2}}\right)^{-3 (w+1)} \,,
\end{eqnarray}
where we have used Eq. (\ref{NN}) to express all in term of the boundary term $B$. The above equation is difficult to solve analytically
for all values of $w$, so that for simplicity we assume the cold dust case $w=0$, which gives us
\begin{eqnarray}\label{rec2}
F(B)&=&+B C_1+\frac{\log \left(B-18 H_0^2\right) \left(6 a_0^3 B l-2 B \kappa ^2 \rho_0\right)+6 a_0^3 l \left(B-9 H_0^2\right)+\kappa
^2
\rho_0 \left(B-36 H_0^2\right)}{27 a_0^3 l}\nonumber\\
&&+\frac{C_2 \left(3 H_0 \sqrt{B-9 H_0^2}-B \arctan\left(\frac{\sqrt{B-9 H_0^2}}{3 H_0}\right)\right)}{54 H_0^3}\,,
\end{eqnarray}
where $C_{1}$ and $C_{2}$ are integration constants.\\
Finally, let us reconstruct the phantom behaviour scenario where the energy density and the Hubble parameter are given by Eq.
(\ref{phantom}). By using that $N=\frac{1}{2m}\ln(B/(6h_{0}^2(3+m)))$, Eq. (\ref{fTB111}) becomes
\begin{eqnarray}
\frac{2 B^2 m F_{BB}(B)}{m+3}-B F_{B}(B)+F(B)+\frac{ B}{m+3}=\nonumber\\
2 \kappa ^2 \left(\frac{B \left(8 \sqrt{6} B^{3/2} b_{2} (m+1)+15 b_{1} h_{0}^3 (m+3)^{3/2}\right)}{90 h_{0}^5
(m+3)^{5/2}}+b_{0}\right)\,,
\end{eqnarray}
and then the corresponding reconstruction function is
\begin{eqnarray}
F(B)&=&-\frac{B \left((m-3) \log (B) \left(3 h_{0}^2-b_{1} \kappa ^2\right)+2 b_{1} \kappa ^2 m-3 h_{0}^2 \left(C_2 m^2-6 C_2 m+9 C_2+4
m\right)\right)}{3 h_{0}^2 (m-3)^2}\nonumber\\
&&+\frac{16 \sqrt{\frac{2}{3}} B^{5/2} b_{2}\kappa ^2 (m+1)}{45 h_{0}^5 (m+3)^{3/2} (4 m-3)}+C_1 B^{\frac{m+3}{2 m}}+2 b_{0} \kappa
^2\,,
\end{eqnarray}
where $C_{1}$ and $C_{2}$ are integration constants. Let us stress here that the final expressions for the function $F(B)$ becomes less complicated that in $f(T,B)$ gravity. For example, in the reconstruction of $\Lambda$CDM, for the $f(T,B)$ model, it was found that the expression contains hypergeometric expressions (see Eq.~\ref{rec1}) where in the $-T+F(B)$ reconstruction (see Eq.~\ref{rec2}), the expression does not have such complicated terms. This is another reason why we studied the reconstruction method in a general setting and then in a specific theory such as $-T+F(B)$ gravity.

\section{Perturbations and Stability}

In this section, we are interested to establish the stability
conditions for cosmological solutions against linear isotropic
homogeneous perturbations in $f(T,B)$ theory of gravity. The perturbation equations in FLRW universe will be formulated for a
general framework and then de-Sitter
and power-law solutions will be studied. We assume a general solution
\begin{equation}\label{s1}
H(t)=H_j(t)\,,
\end{equation}
which satisfies the basic equations of motion of FLRW universe in
$f(T,B)$ theory of gravity. In term of above solution, the torsion
scalar and boundary $B$, can be written as follows
\begin{eqnarray}\label{s2}
T_j&=&6{H}^2_j(t)\,,\\\label{s3} B_j&=&6{\dot{H}}_j(t)+18{{H}^2}_j(t)\,.
\end{eqnarray}
If one considers particular model of $f(T,B)$ that can generate
solution (\ref{s1}), then, the following equations must be satisfied
\begin{eqnarray}\label{s4}
&&-3H^2_j\Big(3f^j_B+2f^j_T\Big)+3H_j\dot{f}^j_B-3\dot{H}_jf^j_B+\frac{1}{2}f^j=\kappa^2
{\rho}_{\rm m}{}_j\,,
\\\label{s5}
&&{\dot{\rho}}_{\rm m}{}_j+3H_j(1+w){\rho}_{\rm m}{}_j=0\,.
\end{eqnarray}
Now we define the perturbation for Hubble parameter and energy
density as follows
\begin{equation}\label{s6}
H(t)=H_j(t)\Big(1+\delta(t)\Big)\,,\quad\quad\quad\quad
\rho_{m}(t)={\rho}_{\rm m}{}_j\Big(1+\delta_{\rm m}(t)\Big)\,.
\end{equation}
Here, our purpose is to make the perturbation analysis about the
solution $H(t)=H_j(t)$, so that function $f(T,B)$ can be expressed
in the powers of $T$ and $B$ as
\begin{equation}\label{s7}
f(T,B)=f^j+f^j_T(T-T_j)+f^j_B(B-B_j)+\mathcal{O}^2\,,
\end{equation}
where the superscript $j$ means the values of $f(T,B)$ and its
derivatives are evaluated at $T=T_j$ and $B=B_j$. The term
$\mathcal{O}^2$ includes all the terms which have power-square and
higher powers of $T$ and $B$, although we shall only consider the
linear terms of the defined perturbation. Thus, by replacing
Eqs.~(\ref{s6}) and (\ref{s7}) in the FLRW equation (\ref{s4}) and in the
continuity equation (\ref{s5}), we get the perturbation equations in terms of
$\delta(t)$ and ${\delta_{\rm m}(t)}$, (in the linear approximation) in
the form of the following differential equations
\begin{eqnarray}\label{s8}
&&c_2\ddot{\delta}(t)+c_1\dot{\delta}(t)+c_0{\delta}(t)=c_{\rm m}
\delta_{\rm m}(t)\,,\\\label{s9}
 &&\dot{\delta_{\rm m}}(t)+3H_j \delta(t)=0\,.
\end{eqnarray}
The coefficients $c_{0,1,2,\rm m}$, are expressed in the Appendix (see \eqref{s100}). These coefficients
depend explicitly on $f(T,B)$ and its derivatives evaluated at
background solutions $H=H_j$. In general it is not easy to solve the
above equations analytically. In the coming sections we shall
present some particular models for the solution of above equations.
\subsection{Stability of de-Sitter Solution}

Consider the de-Sitter solution with $H_j=H_0$ and ${\rho}_0=0$, then the
perturbed equation takes the following form,

\begin{eqnarray}\label{s10}
&&\Big(\ -18{H^2}_0
f^0_{TB}T_0+324H^4_0f^0_{BB}-36H^2_0f^0_B-12H^2_0f^0_{TT}T_0+216H^4_0f^0_{TB}\nonumber\\
&&-24H^2_0f^0_T-24H^2_0f_B\Big)\delta(t)+\Big(-54H^3_0f^0_{BB}-6H_0f^0_{B}\Big)\dot{\delta}(t)+(-18H^2_0f^0_{BB})\ddot{\delta}(t)=0\,.\nonumber\\
\end{eqnarray}
Using the $f(T,B)$ model formulated in de-Sitter reconstruction, i.e., Eq.~(\ref{dS1}), one gets the following solution for $\delta(t)$
\begin{eqnarray}\label{s11}
\delta(t)= C_1e^{{\mu}_{+}t}+C_2e^{{\mu}_{-}t}\,,
\end{eqnarray}
where $C_1$ and $C_2$ are integration constants and
\begin{equation}\label{s12}
{\mu}_{\pm}=\frac{3 H_0}{2 f_0}  \left(-3 f_0\pm\sqrt{{f_0
\left(f_0-28 \sqrt{e} \tilde{f_0}\right)}}\right)\,.
\end{equation}
Here, $f_0$ and $\tilde{f_0}$ are the constants appearing in
Eq.~(\ref{dS1}). {Note that $\sqrt{e}=e^{1/2}$ is referring to the
exponential $e$ and not the determinant of the tetrad.} The growth
of the perturbation will depend both upon the overall sign of the
parameters ${\mu}_{\pm}$ appearing in the expression (\ref{s12}) and
also upon the real and imaginary character of the square root. Thus
four different cases can be distinguished:

\begin{itemize}
\item $f_0<0$ and $f_0>28\sqrt{e}\tilde{f}_0$ with $\tilde{f}_0<0$,
this implies that solutions are complex and $\Re
({\mu}_{\pm})<0$, thus solutions behave as a damped oscillator of
decreasing amplitude. Hence, solutions are stable.
\item $f_0>0$ and $f_0<28\sqrt{e}\tilde{f}_0$ with $\tilde{f}_0>0$,
this implies that solutions are complex and $\Re ({\mu}_{\pm})<0$,
thus solutions behave as a damped oscillator of decreasing
amplitude. Hence, solutions are stable.
\item $0<28\sqrt{e}\tilde{f}_0/f_0<1$ with $f_0>0$ and $\tilde{f}_0>0$,
or $f_0<0$ and $\tilde{f}_0<0$, then both ${\mu}_{\pm}$ are real and
${\mu}_{\pm}<0$, hence solutions are stable.

\end{itemize}
\subsection{Stability of Power Law  Solutions}
\begin{figure}[H]
\center\epsfig{file=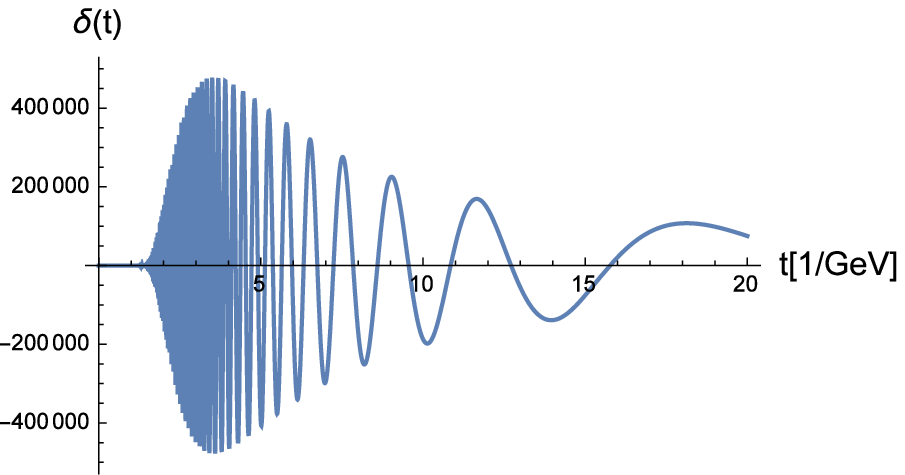, width=0.45\linewidth}\epsfig{file=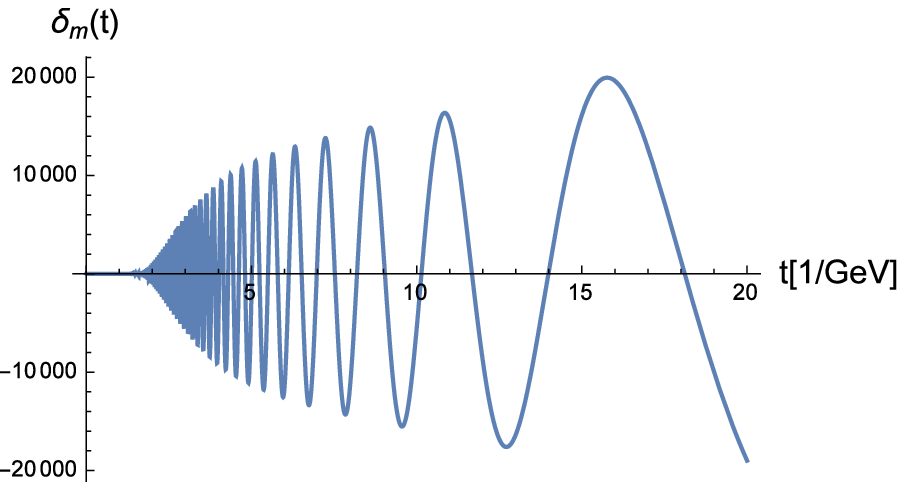, width=0.45\linewidth}
\caption{Evolution of $\delta(t)$ and $\delta_{\rm m}(t)$ versus time $t$. Herein,
we set the initial conditions $\delta'(1)=0.2$, $\delta(1)=0.1$ and $\delta_{\rm m}(1)=0.1$. The figures show the evolution of perturbation
parameters $\delta(t)$ and $\delta_{\rm m}(t)$ for the matter dominated solutions.}
\label{fig44}
\end{figure}

\begin{figure}[H]
\center\epsfig{file=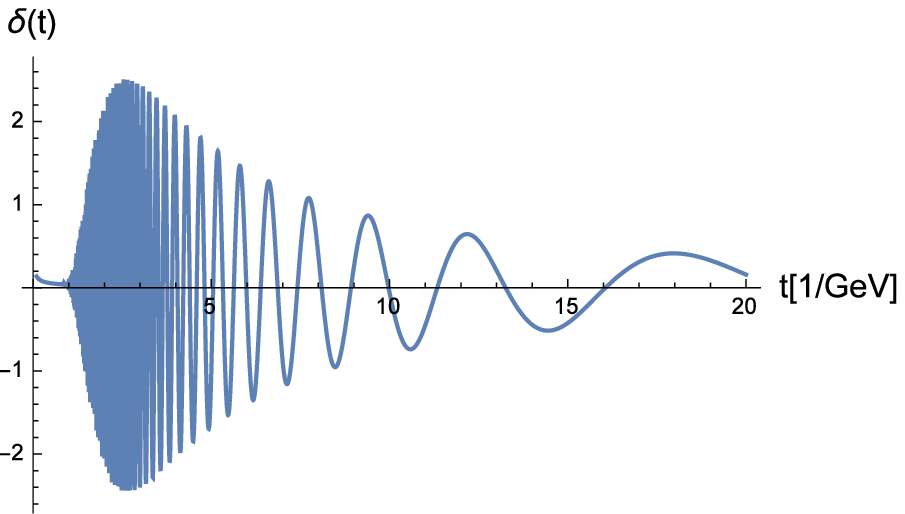, width=0.45\linewidth}\epsfig{file=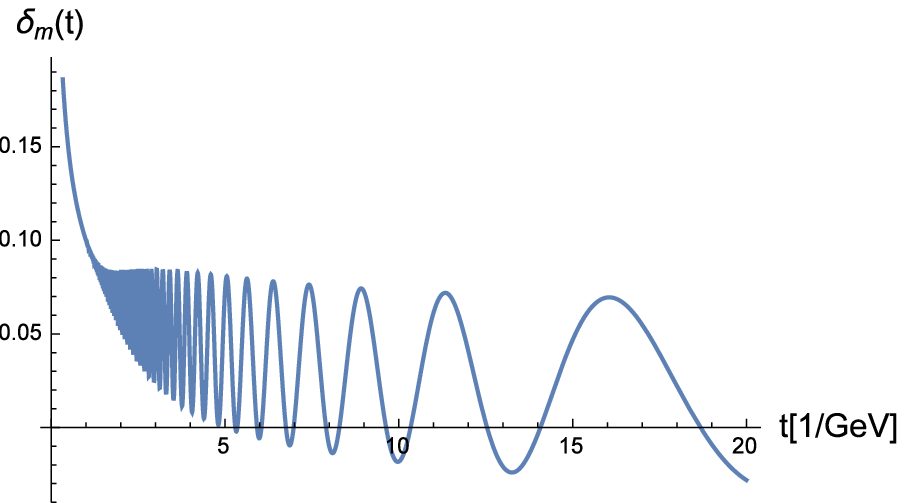, width=0.45\linewidth}
\caption{Evolution of $\delta(t)$ and $\delta_{\rm m}(t)$ versus time $t$. Herein,
we set the initial conditions $\delta'(1)=0.2$, $\delta(1)=0.1$ and $\delta_{\rm m}(1)=0.1$. This Figure shows the evolution of perturbation
parameters $\delta(t)$ and $\delta_{\rm m}(t)$ for the radiation dominated solutions.}
\label{fig55}
\end{figure}

\begin{figure}[H]
\center\epsfig{file=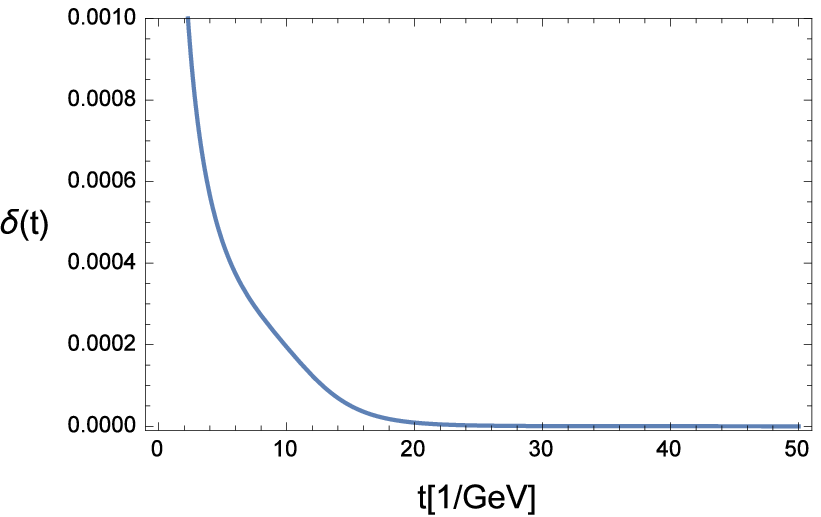, width=0.45\linewidth}\epsfig{file=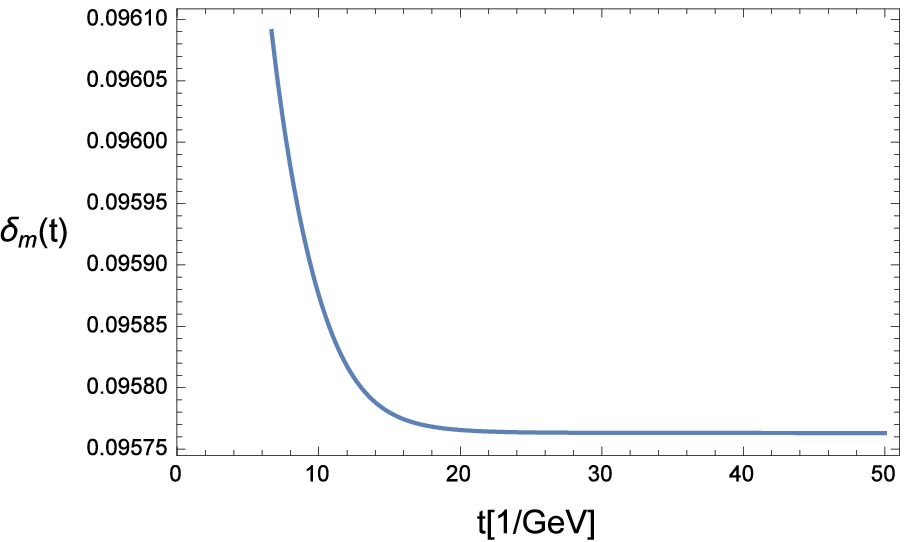, width=0.45\linewidth}
\caption{Evolution of $\delta(t)$ and $\delta_{\rm m}(t)$ versus time $t$. Herein,
we set the initial conditions $\delta'(1)=0.2$, $\delta(1)=0.1$ and $\delta_{\rm m}(1)=0.1$. This Figure shows the evolution of perturbation
parameters $\delta(t)$ and $\delta_{\rm m}(t)$ for the accelerated expansion solutions.}\label{fig66}
\end{figure}

In this section, the stability of the power-law solution described in~(\ref{42}) will be studied. For $0<h<1$, we have decelerated universe
which may refer to dust dominated ($h=2/3$) or
radiation dominated ($h=1/2$), while $h>1$ results
in accelerating picture of the universe.
Here, we explore the stability of power law solutions for
matter dominated, radiation dominated and late time accelerated eras.
\begin{itemize}
  \item For matter dominated era with $h=2/3$, and $w=0$, Eqs.~(\ref{pfTB}) and (\ref{pfTB1})
  result in
  \begin{equation}\label{MD}
  f(T, B)=\frac{C_2}{\sqrt{B}}+C_3B+C_1\sqrt{T}-\frac{3}{4}\rho_{0}\kappa^2 T\,.
  \end{equation}
 By substituting the above model in equations (\ref{s8}) and (\ref{s9}), one can find the required perturbation equations for matter
 dominated a power-law model. Here, we employ the numerical approach to solve these equations and present the evolution of
 perturbation
 parameters $\delta(t)$ and $\delta_{\rm m}(t)$. In this study we set $H_0=67.3$, $\Omega_{\rm m}=0.23$, $C_2=-0.2$, $C_1=C_3=0.1$ and
 $\kappa^2=1$. Fig.~\ref{fig44} shows the oscillating behavior of $\delta(t)$ and $\delta_{\rm m}(t)$, however these do not decay in future
 evolution.
\end{itemize}
\begin{itemize}
  \item For radiation dominated era with $h=1/2$, and $w=1/3$, Eqs.~(\ref{pfTB}) and (\ref{pfTB1})
  result in
  \begin{equation}\label{RD}
  f(T, B)=\frac{C_2}{B^{\frac{1}{4}}}+C_3B+C_1\sqrt{T}-\frac{4}{3}\kappa^2\rho_{0} T\,.
  \end{equation}
  One can substitute the model (\ref{RD}) in Eqs.~(\ref{s8}) and (\ref{s9}) to find the required
   perturbation equations for radiation dominated power law model.
 The numerical scheme and the evolution of $\delta(t)$ and $\delta_{\rm m}(t)$ is depicted in Fig.~\ref{fig55}. The figures show an oscillating behavior of $\delta(t)$ and
  $\delta_{\rm m}(t)$, however the oscillations of $\delta(t)$ and $\delta_{\rm m}(t)$ do not decay in future. Hence
  solutions are unstable as full perturbation around a cosmological solution is
fully determined by the matter perturbations. This result is similar
to matter dominated era with $h=2/3$, and $w=0$ which is shown
in Fig.~\ref{fig44}.
\end{itemize}
\begin{itemize}
  \item For the choice of $h>1$, the universe is expanding. In our case, we set $h=2$ with $w=-0.5$, so that the
      corresponding power law model is given by
  \begin{equation}\label{AEX}
  f(T, B)=\frac{C_2}{B^{\frac{5}{2}}}+C_3B+C_1\sqrt{T}-\frac{\kappa^2\rho_{0} T^{\frac{3}{2}}}{48\sqrt{6}}\,.
  \end{equation}
  Again following a similar approach, we show the results in Fig.~\ref{fig66}.
  Here, we set  $C_1=0.1$ and $C_2=-10,~C_3=-1$. For this case it can be
  seen that $\delta(t)$ and $\delta_{\rm m}(t)$ decay in later times so that
  power-law model (\ref{AEX}) for $h>1$ is stable to some extent.
\end{itemize}

\section{Conclusions}

Over the last years, teleparallel theories of gravity and its modifications
have attained significant attention to address various issues in cosmology.
These theories lie in a globally flat manifold endorsed with torsion. It is
well-known that GR has an equivalent teleparallel representation (TEGR) based
on the torsion (and tetrads) instead of curvature (and metric). In this
perspective, different modified teleparallel theories have been proposed. The
first one, is the so-called, $f(T)$ gravity, a natural generalisation of the
TEGR action by changing the torsion scalar $T\rightarrow f(T)$ in the action.
This approach is analogous with $f(R)$ gravity in the metric counterpart.
These two theories have been very successful describing the cosmological
behaviour of the universe. With the aim to unify both $f(R)$ and $f(T)$
gravity and see how these theories are connected, it was formulated a
modified teleparallel theory of gravity named as $f(T,B)$ theory which under
suitable limits can recover $f(T)$ or $f(R)$ gravity \cite{13*}. In this work,
we have explored different cosmological features in $f(T,B)$ gravity as the
establishment of laws of thermodynamics, reconstruction of some cosmological
models and stability of some models corresponding to linear homogeneous
perturbations.

In Sec.~III, we have shown that the modified flat FLRW equations in
this theory can be cast to the form of first law of thermodynamics,
$\tilde{T}_hdS_h+\tilde{T}_{h}d\bar{S}=-dE+WdV$. Here, $d\bar{S}$ is the additional entropy
term due to non-equilibrium thermodynamics which may be produced as a result
of Lagrangian dependence both on the torsion scalar and the boundary term.
The entropy production term in $f(T,B)$ gravity is more general and can
reproduce the corresponding factor in $f(-T+B)=f(R)$ \cite{44} and $f(T)$
\cite{45} theories. It is worth mentioning that no such term is present in
GR, Gauss-Bonnet gravity \cite{32}, Lovelock gravity \cite{35,7*} and
braneworld gravity \cite{36,37}. Moreover, in case of $f(R)$, $f(T)$ and
scalar tensor theories different schemes have been suggested to avoid the
auxiliary term in first law of thermodynamics \cite{45,ET}. Bamba et al.
\cite{45} show that one can redefine the energy momentum tensor contributed
from the modified theories so that the conservation equation is truly
satisfied and hence results in omission of entropy production term. In case
of $f(T,B)$ gravity we find that one can establish the equilibrium
description of thermodynamics (presented in Sec.~III-B) and remove the additional entropy production
term. We also establish the GSLT which is found to be valid for the phantom
era of cosmos.

We reconstructed the gravitational action of this model and have done a brief
analysis of validity of GSLT and stability of reconstructed models. Here, we
have used a more comprehensive approach for cosmological reconstruction of $f
(T,B)$ gravity in terms of e-folding representing different eras of the
universe. We have studied some important cosmological solutions in the
standard cosmological concordance model around a spatially flat FLRW
background namely dS expansion, power laws and the scale factor solutions as
provided for the $\Lambda$CDM model and phantom dominated model. One can
employ the reconstructed $f(T,B)$ to explore cosmic evolution in more
consistent way. Additionally we have explored the specific case $f(T,B)-T+F(B)$,
which is a GR background plus an additional function that depends on the boundary term.
The reconstruction scheme is carried out for this specific case and we have also obtained
the corresponding function $F(B)$ which mimics different cosmological models. One can employ the
reconstructed $f(T,B)$ models to explore cosmic evolution in more consistent way.

We also examined the validity of GSLT: for the power solutions we find the
validity constraints in case of specific model (\ref{pfTB})-(\ref{pfTB1})
which predicts a late-time accelerating universe. In this case we restrict
the values of integration constant $C_1\geq0$ and vary values of $h$ to see evolution of GSLT. In case of de-Sitter
model, GSLT is trivially satisfied and for $\Lambda$CDM reconstruction
(\ref{z*})-(\ref{z**}), one needs to fix $C_3=0.1$. In case of phantom dominated model, GSLT is valid for
negative values of parameters $b_1$ and $b_2$ (see Eq.~(\ref{phantom}))
together with all values of other parameters $h_0$, $m$ and
$C_3$. The study of stability/instability of various forms of Lagrangian is a
useful tool to classify the modified theories on physical grounds. The
linearised perturbed equations were derived by implementing the perturbation
for the Hubble parameter and energy density. We analyzed the stability of de-Sitter and power law solutions, finding that de-Sitter model is found to be stable
with some constraints on model parameters whereas as power law solution is
stable only for $h>1$ representing expanding behavior of
universe. Hence, we conclude that power law solution is found to be more
feasible as it validates GSLT and is stable against homogeneous perturbation.

It is stated that stability of linear homogeneous perturbations does not
guarantee the stability of the reconstructed $f(T,B)$ models. In future
project we will develop complete set of differential equations for the matter
density perturbations and analyze the growth index to constrain the viable
models.

\renewcommand{\theequation}{A.\arabic{equation}}
\setcounter{equation}{0}
\section*{Appendix A}\label{appendixx}

\begin{eqnarray}\label{s100}
c_0&&=\Big(\ -18{H^2}_j
f^j_{TB}T_j+324H^4_jf^j_{BB}+54H^2_j\dot{H}_jf^j_{BB} -36H^2_jf^j_B
-12H^2_jf^j_{TT}T_j\nonumber\\
&&+216H^4_jf^j_{TB}+36H^2_j\dot{H}_jf^j_{TB}-24H^2_jf^j_T+6H_j\dot{f}^j_{TB}T_j+6H_j{f}^j_{TB}\dot{T}_j
-108H^3_j\dot{f}^j_{BB}\nonumber\\
&&-18\dot{H}_jH_j{\dot{f}}^j_{BB} -216H^2_j\dot{H}_jf^j_{BB}
-18H_j\ddot{H_j}f^j_{BB}+6H_j\dot{f^j}_B-6\dot{H}_jT_jf^j_{TB}\nonumber\\
&&+108\dot{H}_jH^2_jf^j_{BB}+18{\dot{H}}^2_jf^j_{BB}
-9\dot{H}_jf^j_{B}+f^j_BT_j-18f^j_B{H_j}^2-3f^j_B\dot{H}_j\Big)\\
c_1&&=54H^3_jf^j_{BB}+36H^3_jf^j_{TB}+6H_jf^j_{TB}T_j-18H^2_j\dot{f}^j_{BB}-108H^3_jf^j_{BB}\nonumber\\
&&-6H_jf^j_B\,,\\
c_2&&=-18{H_j}^2f^j_{BB}\,,\\
c_{\rm m}&&=\kappa^2 \rho_{\rm m}\,.
\end{eqnarray}

\begin{flushleft}
    \textbf{Acknowledgments}
\end{flushleft}
S.B. is supported by the Comisi{\'o}n Nacional de Investigaci{\'o}n
Cient{\'{\i}}fica y Tecnol{\'o}gica (Becas Chile Grant No.~72150066). M.
Zubair thanks the Higher Education Commission, Islamabad, Pakistan for its
financial support under the NRPU project with grant number
$\text{5329/Federal/NRPU/R\&D/HEC/2016}$.

\end{document}